\documentclass{aa}  

\usepackage{graphicx}
\usepackage{txfonts}
\usepackage{hyperref}
\usepackage{siunitx}
\usepackage{color}

\usepackage{amsmath}

\def\reff{R_{\mathrm{e}}}
\def\lowre{r_{\mathrm{e}}}
\def\mstar{M_*}

\def\mlowstarobs{m_*^{(\mathrm{obs})}}

\def\zsource{z_{\mathrm{s}}}
\def\zlens{z_{\mathrm{g}}}

\def\mfive{M_{5}}
\def\mfiveobs{m_{5}^{(\mathrm{obs})}}

\def\tein{\theta_{\mathrm{E}}}
\def\teinest{\theta_{\mathrm{E}}^{(\mathrm{est})}}
\def\teinsis{\theta_{\mathrm{E}}^{(\mathrm{SIS})}}
\def\teinobs{\theta_{\mathrm{E}}^{(\mathrm{obs})}}

\def\hyperpars{\boldsymbol{\eta}}

\def\psilens{\boldsymbol{\psi}_\mathrm{g}}
\def\psilensi{\boldsymbol{\psi}_{\mathrm{g},i}}
\def\psisource{\boldsymbol{\psi}_\mathrm{s}}
\def\psisourcei{\boldsymbol{\psi}_{\mathrm{s},i}}

\def\sourcepos{\boldsymbol\beta}

\def\sourcelight{\boldsymbol{\psi}_{\mathrm{s}}^{\mathrm{l}}}

\def\prlens{{\rm P}_\mathrm{g}}
\def\prsource{{\rm P}_\mathrm{s}}
\def\prsourceeff{{\rm P}_\mathrm{s}^{(\mathrm{eff})}}

\def\prsl{{\rm P}_\mathrm{{SL}}}
\def\prsleff{{\rm P}_\mathrm{{SL}}^{(\mathrm{eff})}}

\def\pdet{{\rm P}_\mathrm{det}}
\def\psel{{\rm P}_\mathrm{sel}}
\def\pfind{{\rm P}_\mathrm{find}}

\def\crosssect{\sigma_\mathrm{{SL}}}

\def\crosssectbosix{\sigma_\mathrm{{SL}}^{(\mathrm{B06})}}

\def\sigmaap{\sigma_\mathrm{{ap}}}
\def\sigmaapobs{\sigma_\mathrm{{ap}}^{(\mathrm{obs})}}

\def\sigmafp{\sigma_{\mathrm{FP}}}
\def\slbias{b_{\mathrm{SL}}}

\def\data{\mathbf{d}}
\def\datai{\mathbf{d}_i}

\def\datagal{\mathbf{d}_{\mathrm{g}}}

\def\Sref#1{Section~\ref{#1}\xspace}
\def\Fref#1{Figure~\ref{#1}\xspace}
\def\Tref#1{Table~\ref{#1}\xspace}
\def\Eref#1{Equation~\ref{#1}\xspace}

\def\pr{{\rm P}}

\setcitestyle{notesep={}}

\begin{document}

   \title{The SLACS strong lens sample, debiased}
   \titlerunning{SLACS, debiased}
   \authorrunning{Sonnenfeld}

   \author{Alessandro Sonnenfeld\inst{\ref{sjtu1},\ref{sjtu2},\ref{sjtu3}}
          }

   \institute{
Department of Astronomy, School of Physics and Astronomy, Shanghai Jiao Tong University, Shanghai 200240, China\\
              \email{sonnenfeld@sjtu.edu.cn}\label{sjtu1} \and
Shanghai Key Laboratory for Particle Physics and Cosmology, Shanghai Jiao Tong University, Shanghai 200240, China\label{sjtu2} \and
Key Laboratory for Particle Physics, Astrophysics and Cosmology, Ministry of Education, Shanghai Jiao Tong University, Shanghai 200240, China\label{sjtu3}
             }

   \date{}

 
  \abstract
    {
Strong gravitational lensing observations can provide extremely valuable information on the structure of galaxies, but their interpretation is made difficult by selection effects, which, if not accounted for, introduce a bias between the properties of strong lens galaxies and those of the general population.
A rigorous treatment of the strong lensing bias requires, in principle, to fully forward model the lens selection process.
However, doing so for existing lens surveys is prohibitively difficult.
With this work we propose a practical solution to the problem: 
using an empirical model to capture the most complex aspects of the lens finding process, and constraining it directly from the data together with the properties of the lens population. 
We applied this method to real data from the SLACS sample of strong lenses.
Assuming a power-law density profile, we recovered the mass distribution of the parent population of galaxies from which the SLACS lenses were drawn.
We found that early-type galaxies with a stellar mass of $\log{\mstar/M_\odot}=11.3$ and average size have a median projected mass enclosed within a $5$~kpc aperture of $\log{\mfive/M_\odot}=11.332\pm0.013$, and an average logarithmic density slope of $\gamma=1.99\pm0.03$.
These values are respectively $0.02$~dex and $0.1$ lower than inferred when ignoring selection effects.
According to our model, most of the bias is due to the prioritisation of SLACS follow-up observations based on the measured velocity dispersion.
As a result, the strong lensing bias in $\gamma$ reduces to $\sim0.01$ when controlling for stellar velocity dispersion.
}

   \maketitle
%

\section{Introduction}\label{sect:intro}

Strong gravitational lensing is a unique tool for measuring galaxy masses at cosmological distances. 
Strong lensing constraints have enabled us to gain insight on the mass structure of the most massive galaxies, as well as on cosmology \citep[see][ and references therein]{Sha++24}.
Strong lenses, however, are a special subset of the general galaxy population, for two reasons.
First, the probability of a galaxy to act as a strong lens varies as a function of its properties: for example, the more massive and more concentrated a galaxy is, the more likely it is to be a lens.
Second, lenses of different properties have different probabilities of being discovered by a strong lensing survey: in general, lenses with a larger Einstein radius (and hence a larger mass, everything else being fixed) are more likely to be identified \citep[see e.g.][]{HOV23}.
The ratio between the probability distribution of galaxy properties that is sampled by a strong lens survey and that of the parent population from which the lenses are drawn is referred to as the strong lensing bias.
This bias, if not accounted for, makes it difficult to compare strong lensing measurements to other measurements or models.

One possible approach to mitigate the strong lensing bias is to control for observed galaxy properties when comparing lenses with non-lens galaxies.
For example, the stellar mass distribution of strong lens galaxies is very different from that of the rest of the population. 
However, if a given property X correlates with stellar mass, differences in X between the lens and non-lens sample will be reduced when comparing them at fixed stellar mass.

In practice, even at fixed stellar mass strong lens galaxies are noticeably different from non-lens galaxies: for example, they are on average more compact \citep[see e.g.][]{Aug++10}.
Controlling for central velocity dispersion appears to be a better choice: 
 lens galaxies from the Sloan Lens ACS Survey \citep[SLACS][]{Bol++06} have been shown to be indistinguishable from non-lens galaxies in terms of their fundamental plane\footnote{The fundamental plane is a scaling relation between half-light radius, central surface brightness and central stellar velocity dispersion of early-type galaxies \citep{Dre++87,D+D87}.} \citep{Tre++06} and environment \citep{Tre++09}, at fixed velocity dispersion.
This can be explained by noting that, under the assumption of an isothermal density profile, the probability of a galaxy to create multiple lensed images of a background source is a function of velocity dispersion alone. Thus, to first approximation, strong lens samples can be viewed as being velocity dispersion-selected. However, galaxies are not exactly isothermal systems, and the selection of lenses involves more criteria than the simple production of multiple images (as we discuss below). Therefore we expect that not all of the strong lensing bias can be captured by velocity dispersion alone.

In general, finding analogues to galaxies acting as strong lenses is an ill-posed problem, because the strong lensing bias depends on the mass structure of galaxies, which is difficult to know exactly (and usually is the very property that one seeks to constrain with strong lensing).
Instead, a more productive approach to dealing with the strong lensing bias is to forward model the selection process of lenses in a survey.
\citet{Son22} showed an example of how this can be done when working with a complete sample of lenses and well-defined selection criteria.
For most of the strong lensing surveys carried out so far, however, it is very challenging to model the selection function from first principles.
This is because existing lens samples are partly the result of human decisions on the classification of lens candidates, which are difficult to reproduce and include in a model.

In this paper we propose a practical solution to the problem of correcting for the strong lensing bias in the analysis of a strong lens sample.
We tackled the problem by making some simplifying assumptions on the lens detection probability, and by using empirical models to describe those aspects of the problem that are most difficult to predict.
These empirical models are to be inferred from the data along with the mass properties of the population of galaxies.
We first explain our methodology in general terms, then show its application to the SLACS strong lens sample.

We used our technique to revisit measurements of the total density profile of massive galaxies.
Strong lensing and stellar dynamics observations based on the SLACS sample have revealed how, under the assumption of a power-law density profile, $\rho(x)\propto x^{-\gamma}$, the average density slope of early-type galaxies is $\left<\gamma\right> = 2.09$, corresponding to a distribution slightly steeper than isothermal, and scales positively with stellar surface mass density \citep{Koo++06,Aug++10,Son++13b}.
The trend with stellar surface mass density can be easily explained with the dark matter halo contributing to a smaller fraction of the total mass within the half-light radius in more compact galaxies \citep{Sha++17}.
However, when combining SLACS with other lens samples at higher redshift, such as the BOSS Emission-Line Lens Survey \citep[BELLS][]{Bol++12} or the Strong Lens Legacy Survey \citep[SL2S][]{Gav++12}, lensing and dynamics constraints indicate that the average slope $\gamma$ increases from $z=1$ to the present \citep{Ruf++11, Bol++12, Son++13b}, in a way that is difficult to reconcile with theory. Hydrodynamical simulations and empirical models tend to predict opposite trends \citep{Rem++13, Xu++17, Sha++18}, with the exception of {\sc HORIZON-AGN} \citep{Pei++19}.
The issue of the evolution in $\gamma$ is related to the late growth history of massive galaxies: these objects are generally believed to grow as a result of dissipationless mergers, which tend to make the density slope shallower \citep{SNT14}. 
One possible way to reconcile theory and observations is then to allow for some amount of dissipation and related gas infall associated with mergers \citep{SNT14}.
As an alternative to the dissipational merger scenario, it has been suggested that strong lensing selection effects are responsible for the observed trend \citep[e.g.][]{Rem++13}.
With this work we aim to bring new insight to the problem, by re-analysing the SLACS sample while accounting for the strong lensing bias.

The structure of this paper is as follows.
In \Sref{sect:sample} we describe the lens sample that we use for our case study, with a focus on the lens selection process.
In \Sref{sect:method} we introduce a formal definition of the strong lensing bias and describe the general methodology that we adopt to account for it.
In \Sref{sect:model} we describe the model that we fit to the lensing data.
In \Sref{sect:results} we show the results of our analysis.
We discuss the results in Sect. \ref{sect:discuss} and draw conclusions in Sect. \ref{sect:concl}.

The Python code used for the simulation and analysis of the lens sample can be found in a dedicated section of a GitHub repository\footnote{\url{https://github.com/astrosonnen/strong_lensing_tools/tree/main/papers/slacs_selection}}.
Throughout this work we assumed a flat $\Lambda$ cold dark matter model, with $H_0 = 70\,\mathrm{km}\,\mathrm{s}^{-1}\,\mathrm{Mpc}^{-1}$, and $\Omega_m = 0.3$.


\section{The strong lens sample}\label{sect:sample}

Our study focuses on lenses drawn from the SLACS survey.
In this section we first describe the selection process of these lenses, and then their associated data.

\subsection{The selection process}\label{ssec:slacssel}

SLACS lenses were discovered as follows.
Foreground galaxies were selected among the main and the luminous red galaxy (LRG) spectroscopic samples \citep{Eis++01,Str++02} of the Sloan Digital Sky Survey \citep[SDSS][]{Yor++00}.
The SDSS spectra of the foreground galaxies were scanned for the presence of emission lines associated with objects at higher redshift.
Among the systems with a spectroscopic detection of a background galaxy, those expected to have the largest strong lensing cross-section 
(we explain below how it was estimated) 
were followed-up with the Hubble Space Telescope (HST) for high-resolution imaging \citep{Bol++06}.
Lenses with a clear detection of a set of strongly lensed arcs were then modelled for the purpose of measuring their Einstein radius.
 
The resulting selection function of the SLACS sample is rather complex.
In order for the source redshift to be measured, multiple emission lines had to be detected. These lines are typically [OIII], H$\beta$ and the [OII] doublet. Since the wavelength coverage of the SDSS spectrograph extends only up to $\sim9200\AA$, only sources up to redshift $\sim0.9$ could effectively be discovered.
At fixed source redshift, the lens detection probability also depends on the source emission line strength and surface brightness distribution.
Additionally, the finite size of the SDSS spectroscopic fibre, $1.5''$ in radius, put constrains on the Einstein radius and image configuration of the lens. For instance, lenses with a very large Einstein radius have a main arc that falls outside of the fibre, and could only be detected if one of the counter-images is sufficiently close to the centre, or if sufficient flux is brought into the fibre by the point spread function (PSF).
We refer to the simulation study of \citet{Arn++12} for more details on fibre size effects. 

Finally, the prioritisation in terms of lensing cross-section applied by \citet{Bol++06} introduced another non-trivial selection effect.
\citet{Bol++06} estimated the lensing cross-section of each lens candidate as follows. 
First, they assumed a singular isothermal density profile for each lens.
Then, they used the observed stellar velocity dispersion measured by the SDSS, $\sigmaapobs$, as a proxy for the velocity dispersion of the singular isothermal mass distribution. Under this assumption, they estimated the Einstein radius of a lens as
\begin{equation}\label{eq:teinest}
\teinest = 4\pi\left(\frac{\sigmaapobs}{c}\right)^2\frac{D_{\mathrm{ds}}}{D_{\mathrm{s}}},
\end{equation}
where $c$ is the speed of light, $D_{\mathrm{ds}}$ is the angular diameter distance between the lens and the source, and $D_{\mathrm{s}}$ is the angular diameter distance between the observer and the source.
Finally, they defined the strong lensing cross-section as the area in the source plane that is mapped into sets of multiple images\footnote{This definition differs from the one that we adopt in our paper (which we introduce in section \ref{ssec:gterm}).}, which for a singular isothermal sphere is equal to $\pi\tein^2$. 
We indicate this quantity as $\crosssectbosix$.
In summary, $\crosssectbosix$ is a combination of observed stellar velocity dispersion, lens redshift and source redshift.
The way that $\crosssectbosix$ enters the SLACS selection function, however, is rather subtle: 
it was used to rank candidates identified via spectroscopy for the purpose of scheduling follow-up HST observations.
This means that the probability of a lens being included in the SLACS sample depended on the size of the parent sample of lens candidates and on the number of systems with HST data: in the limit of infinite HST time, all candidates would have been followed-up. 
Explicitly incorporating these aspects in a model of the SLACS selection function can be cumbersome.
Moreover, a selection based on the observed velocity dispersion, as opposed to its true value (which is not directly accessible), can lead to the velocity dispersion of the sample to be systematically overestimated. 

In principle, the selection function of the SLACS sample is also determined by the finite resolution of the photometric data:
lenses with an image separation that is close to the size of the PSF may not be recognised as such.
In the case of the SLACS sample, however, we argue that resolution does not play a primary role.
This is because the smallest Einstein radius in the SLACS sample is $\tein=0.79''$, much larger than the FWHM of HST photometric data.
Indeed, the same search methodology at the base of the SLACS selection has been used by \citet{Shu++17} to discover lenses with significantly smaller Einstein radii, down to $\tein\approx0.5''$. 
\citet{Shu++17} used a different criterion to prioritise spectroscopically detected candidates for HST follow-up, compared to SLACS.
We then concluded that the main source of selection on the Einstein radius distribution of the SLACS sample is the lensing cross-section prioritisation, which depends on the observed velocity dispersion.
In \Sref{sect:method} we explain how we can model the overall selection function of the SLACS sample.

\subsection{Sample definition and data}

We took 59 lenses from the SLACS survey, for our study.
These are the systems for which \citet{Aug++10} carried out a joint lensing and dynamical analysis, measuring their mass density slope.
All of the lenses were selected to be early-type galaxies: although the SLACS campaign also led to the discovery of late-type lenses \citep{Tre++11}, these were not included in the \citet{Aug++10} study and are not be considered here either.
The lenses span the redshift range $0.06 < z < 0.36$, with a median redshift of $0.19$.
For the purposes of our analysis, we needed measurements of the lens redshift, Einstein radius, stellar mass, half-light radius and velocity dispersion, as well as the redshift of the source. 
We took these data from \citet{Aug++09}. Einstein radii were measured by fitting singular isothermal ellipsoid models to HST images of the lenses. 
Stellar masses were obtained by fitting stellar population synthesis models to broad band photometric measurements.
\citet{Aug++09} provided measurements of the stellar mass under the assumption of a Chabrier or a Salpeter stellar initial mass function (IMF).
We used Chabrier stellar masses for our analysis (though all of our results can be converted to a Salpeter IMF by rescaling stellar masses upwards by $0.25$~dex).
Rest-frame $V$-band half-light radii were obtained by \citet{Aug++09} by fitting a de Vaucouleurs profile to the HST images.
\Fref{fig:slacsparent} (grey dots) show the distribution in redshift, observed stellar mass, half-light radius and stellar velocity dispersion of the SLACS lenses.

\subsection{The parent population}

As we explained in section \ref{ssec:slacssel}, the first step in the selection of SLACS lenses consisted in scanning SDSS spectra of a large sample of early-type galaxies\footnote{To be precise, the morphology selection was only applied after the lens search, but it is convenient to consider it as part of the initial cut.}.
We refer to that sample as the parent population of foreground galaxies, or simply the parent sample.
The goals of our are analysis are to 1) understand the strong lensing bias of SLACS with respect to the parent population, and 2) recover the mass structure of the parent population galaxies.
To achieve both goals, it is necessary to obtain an accurate description of the parent sample.
Obtaining the exact sample that was used originally for the SLACS search is nearly impossible, 
because \citet{Bol++06} carried out the search in internal data releases of the SDSS, which may not correspond to any of the public data releases available today. 
For this reason, we built our fiducial SLACS parent sample on the SDSS data release 18 \citep[DR18][]{Alm++23} instead, under the assumption that it samples the same galaxy population studied by \citet{Bol++06}.
Starting from the main and LRG spectroscopic samples, we first applied a redshift cut, limiting the sample to galaxies in the range $0.05 < z < 0.40$.
Then, in order to select early-type galaxies, we removed objects with H$\alpha$ emission, by applying an upper limit of $0\AA$ to the H$\alpha$ equivalent width (positive values corresponding to emission).
Finally, we removed galaxies with blue colour. For this purpose, we used the rest-frame absolute magnitudes provided by the Granada Flexible Stellar Population Synthesis \citep[FSPS][]{Con++09} value-added catalogue of DR18 \citep{Mon++16}, which are K- and E-corrected to $z=0.55$, and imposed a minimum $u-r$ value of $2.0$. The resulting sample consists of \num{401576} galaxies.

Given this sample, our inference method required us to obtain a model for their distribution in redshift, stellar mass and half-light radius space.
We obtained stellar mass measurements from the Granada FSPS catalogue, using the masses based on the early star-formation with dust model.
The Granada FSPS masses for the SLACS lenses are on average $0.23$~dex larger than the measurements of \citet{Aug++09}, most likely as a result of a different assumption on the stellar IMF. In order to make these measurements consistent, we applied a $-0.23$~dex shift to the Granada FSPS masses of the entire parent sample.

Finally, we relied on a mass-size relation from the literature to describe the distribution in half-light radius of the parent sample.
In particular, we used the quadratic fit of \citet{H+B09}, which describes the mass-size relation of early-type galaxies over the full stellar mass range of our parent sample.
We also took half-light radius measurements directly from the SDSS, using the $r$-band de Vaucouleurs model from DR18. However, 
these can suffer from contamination from neighbouring objects and deblending issues \citep[see e.g.][]{MVB15}, and therefore cannot be used directly for precision measurements without significant additional work.
We used these half-light radius measurements for illustrative purposes only.

\begin{figure*}
\includegraphics[width=\textwidth]{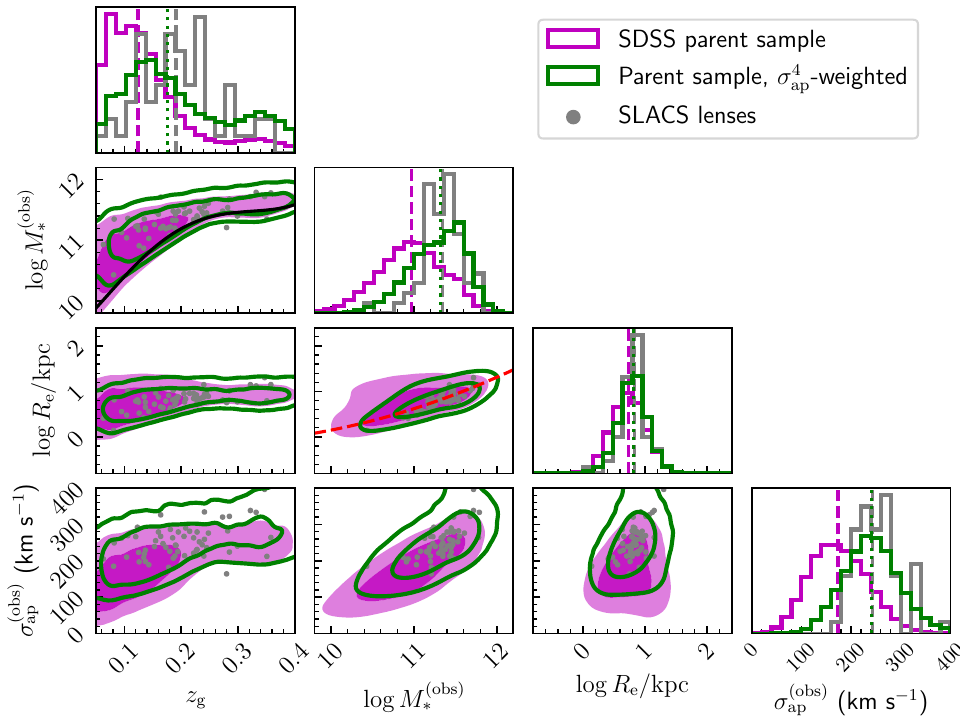}
\caption{
SLACS lenses and parent population.
Distribution in redshift, observed stellar mass, half-light radius, and stellar velocity dispersion.
The green contours and histograms are obtained by weighting the parent sample distribution by $\sigmaap^4$.
Vertical dashed or dotted lines indicate the median of the distributions of the corresponding colour.
Contours enclose 68\% and 95\% of the distribution.
Half-light radius measurements of the parent population are not used in our analysis. Instead, we assumed a stellar mass-size relation from \citet{H+B09}, which is shown as a dashed red line.
The black line in the redshift-stellar mass panel is the inferred truncation stellar mass $m_t$, introduced in section \ref{ssec:gpop}.
\label{fig:slacsparent}
}
\end{figure*}

\Fref{fig:slacsparent} shows the distribution of the parent sample in redshift, stellar mass, half-light radius and stellar velocity dispersion, along with that of the SLACS lenses.
Stellar velocity dispersion measurements were taken from DR18. 
In \Sref{sect:intro} we argued that, to first approximation, strong lens samples are selected in velocity dispersion. To verify the validity of this assertion, we also plot in \Fref{fig:slacsparent} the distribution of the parent sample weighted by $\sigmaap^4$, which is proportional to the estimated lensing cross-section of a singular isothermal sphere with velocity dispersion equal to $\sigmaap$. The medians of the four quantities for the $\sigmaap^4$-weighted distribution are remarkably close to those of the SLACS lenses. However, the detailed shapes of the distributions are significantly different. This suggests that a $\sigmaap^4$-weighting can account for large part of the strong lensing bias, but not its entirety.

\section{Methods}\label{sect:method}

In this section we develop a formalism to correct for the strong lensing bias in a strong lensing inference. First we examine the problem in its full complexity, then we apply a few approximations to make it more tractable in practice. 

\subsection{The problem}\label{ssec:problem}

\citet{Son++23} developed a formalism describing how strong lenses differ from their parent galaxy population.
We follow the same notation in this paper.
We considered a population of foreground galaxies and a population of background sources. 
Each foreground galaxy is described by a set of parameters $\psilens$, drawn from a probability distribution $\prlens(\psilens)$. These parameters describe all aspects that are relevant for predicting the lensing properties (i.e. the mass structure) and the selection of lens candidates (e.g. the stellar mass or the redshift).
Each background source is described by parameters $\psisource$, drawn from a probability distribution $\prsource(\psisource)$. These parameters describe all aspects necessary to predict its observed features, when lensed by a foreground object (i.e. its redshift, position, surface brightness distribution and spectrum).
Then the joint probability distribution in $(\psilens,\psisource)$ of a sample of strong lenses from a given survey is
\begin{equation}\label{eq:one}
\prsl(\psilens,\psisource) \propto \prlens(\psilens)\prsource(\psisource)\psel(\psilens,\psisource),
\end{equation}
where $\psel(\psilens,\psisource)$ is the probability for a galaxy-source pair to produce a strong lens that is identified as such and included in the survey.
The term $\psel$, in other words, describes the selection function of a strong lens sample. 

\Eref{eq:one} describes how the distribution in the properties of lens galaxies and lensed sources (the left-hand side of the equation) differ from that of their parent populations ($\prlens$ and $\prsource$ on the right-hand side).
If we are exclusively interested in the foreground galaxies, we can marginalise over the source distribution to obtain
\begin{equation}\label{eq:slbias}
\prsl(\psilens) \propto \prlens(\psilens)\slbias(\psilens),
\end{equation}
where $\slbias$ is the formal definition of the strong lensing bias and is given by
\begin{equation}\label{eq:slbiasdef}
\slbias(\psilens) \equiv B \int d\psisource \prsource(\psisource)\psel(\psilens,\psisource).
\end{equation}
The constant $B$ is set by the condition that $\prsl(\psilens)$ is normalised to unity.
In order to compare a strong lens sample with non-lens galaxies in an accurate way, it is necessary to either invert \Eref{eq:slbias} and obtain $\prlens$, or to multiply the distribution of non-lenses by $\slbias$, each approach requiring good knowledge of the strong lensing bias.
However, $\slbias$ depends on survey-specific details, such as the distribution of background sources (in particular their redshift, as we discuss later) and the lens finding probability.
In other words, two different strong lensing surveys targeting the same population of foreground galaxies will in general have different strong lensing biases.

We can address the problem by modelling the properties of the foreground population of galaxies, of the background sources, and of the lens finding process. We can assume that all of these aspects of the problem can be fully described by a set of parameters $\hyperpars$, and try to infer the values of these parameters from the data of a strong lens survey, $\data$.
The posterior probability distribution of the parameters $\hyperpars$ is
\begin{equation}
\pr(\hyperpars|\data) \propto \pr(\hyperpars)\pr(\data|\hyperpars),
\end{equation}
where $\pr(\hyperpars)$ is the prior probability and $\pr(\data|\hyperpars)$ the likelihood of observing the data given the model.
Assuming that measurements on individual lenses are independent of each other, the likelihood can be factorised as
\begin{equation}\label{eq:factorlike}
\pr(\data|\hyperpars) = \prod_i^{N_{\mathrm{Lens}}}\pr(\datai|\hyperpars),
\end{equation}
where $N_{\mathrm{Lens}}$ is the total number of lenses.

In order to evaluate individual factors in \Eref{eq:factorlike}, it is necessary to introduce parameters describing each system and average over them. 
That allows us to expand the likelihood of an individual lens as
\begin{equation}\label{eq:integral}
\pr(\datai|\hyperpars) = \int d\psilensi d\psisourcei \pr(\datai|\psilensi,\psisourcei)\pr_{\mathrm{SL}}(\psilensi,\psisourcei|\hyperpars).
\end{equation}

Assuming that the lens selection process can be forward modelled completely, then it is possible to compute integrals of the kind of \Eref{eq:integral}, and thus evaluate the likelihood. Unfortunately, that is not the case for virtually all existing lens surveys, including SLACS. In the next section we provide a strategy for substantially simplifying \Eref{eq:integral}.

\subsection{A solution}\label{ssec:approx}

The first step consists of simplifying the data vector $\datai$.
The data can be though of as the sum of source- and lens-related observables. The former are the redshift, emission line flux, and surface brightness distribution of the lensed images; the latter are the redshift, stellar mass distribution and velocity dispersion of the lens galaxy.
First, we dropped the source emission line flux from $\datai$. 
This does not mean that we ignored the spectroscopic detection process: as we explain below, we described it in an approximate way. Rather, we gave up modelling explicitly the spectroscopic selection from first principles.
Second, we compressed the information from the source surface brightness into a model-independent summary observable: the Einstein radius.
This means that, similarly to the spectroscopic selection, also the photometric detection is not fully modelled, but is described in an approximate way.
The data related to a single lens then becomes
\begin{equation}
\datai \equiv \{\zsource,\teinobs,\datagal\},
\end{equation}
where $\datagal$ includes all of the observables related to the lens galaxy. From here onward we omit the subscript $i$ on individual observables and parameters for ease of notation.
These assumptions allowed us to simplify the first factor in the integrand of \Eref{eq:integral}, $\pr(\datai|\psilens,\psisource)$, as now the only source-related parameter on which the data depends explicitly is the redshift.

As the next step, we proceeded to simplify the strong lens population distribution term $\pr_{\mathrm{SL}}$. 
We began by expanding $\psel$ as follows:
\begin{equation}\label{eq:two}
\psel(\psilens,\psisource) = \pdet(\psilens,\psisource)\pfind(\psilens,\psisource).
\end{equation}
In the above equation, $\pdet$ is the probability of a galaxy-source pair to produce a detectable lens in the survey considered. The term $\pfind$ is instead the probability of the lens being found in the survey, given that it is detectable.
The concept of detectability is rather subtle and deserves further explanation.
First of all, we defined a detected strong lens as a system with at least two images of the same source that are detected and spatially resolved in photometric data, and a spectroscopic measurement of the lens and source redshift.
Therefore, our definition of detection covers both photometry and spectroscopy.
Second, we defined a lens as detectable if the survey data and hypothetical follow-up observations would yield a detection according to our definition.
Crucially, $\pdet$ is different from the probability of a lens being actually detected: the latter depends on the availability of data, which in turn is determined by human decisions on which candidates are followed-up with high-resolution imaging.
These human factors are captured by the factor $\pfind$ in \Eref{eq:two}.

The SLACS sample constitutes only a subset of the strong lenses with a spectroscopic detection in SDSS data. Some of those lenses that are not in SLACS would have been detected photometrically if followed-up with HST \citep[many, in fact, have been detected by][]{Shu++17}. The difference between the pool of detectable lenses and the actual SLACS sample is due to the criteria used for the prioritisation of targets for follow-up, which can be described with $\pfind$.

The use of the concept of detectability might appear like a nuisance, but is made necessary by the fact that the SLACS lens finding process cannot be split cleanly into a detection and a search phase.
\Eref{eq:two} instead allowed us to separate the problem into a part that depends purely on data, $\pdet$, and a part that describes human decision processes, $\pfind$.
In the rest of this section, we provide a practical way to describe the factors entering \Eref{eq:two} explicitly.

We began by writing the source parameters as
\begin{equation}
\psisource = (\zsource,\sourcelight,\sourcepos),
\end{equation}
where $\zsource$ is the redshift, $\sourcepos$ the angular position in the source plane and $\sourcelight$ encodes all information on the light distribution (flux, surface brightness profile and spectrum).
We then simplified the probability of finding a detectable lens, $\pfind$,
by making the following approximation:
\begin{equation}\label{eq:approx}
\pfind(\psilens,\psisource) \approx \pfind(\psilens,\zsource).
\end{equation}
That is, we assumed that the lens finding probability does not depend on any source property other than the redshift.
In the case of SLACS, this is a very good approximation: the decision process that led to the follow-up and subsequent discovery of the SLACS lenses was based purely on the measured values of the lens redshift, lens velocity dispersion, and source redshift (source flux and spectrum only enter $\pdet$).

With the approximations introduced so far, \Eref{eq:integral} could be rewritten as
\begin{align}\label{eq:expandintegral}
\pr(\datai|\hyperpars) = \int & d\psilens d\sourcelight \pr(\datagal|\psilens) \pr(\teinobs|\psilens,\zsource)\prlens(\psilens|\hyperpars)\times \nonumber \\
& \prsource(\zsource,\sourcelight|\hyperpars)\pfind(\psilens,\zsource|\hyperpars)\crosssect(\psilens,\zsource,\sourcelight).
\end{align}
In writing \Eref{eq:expandintegral} we used \Eref{eq:two}, we introduced the strong lensing cross-section,
\begin{equation}\label{eq:csdef}
\crosssect(\psilens,\zsource,\sourcelight) \equiv \int d\sourcepos \pdet(\psilens,\psisource),
\end{equation}
we carried out the integral over the source redshift (because it is known spectroscopically),
and we implicitly assumed that the sources are distributed uniformly in the sky.

To further simplify the problem, we assumed that the strong lensing cross-section can be separated into a geometric factor $g$ that depends purely on the lens properties and source redshift, and a factor $l$ that depends only on the source light distribution:
\begin{equation}\label{eq:csapprox}
\crosssect(\psilens,\zsource,\sourcelight) \propto l(\sourcelight)g(\psilens,\zsource).
\end{equation}
The geometric factor $g$ can be thought of as the strong lensing cross-section computed for a reference value of the source light distribution.
The ansatz of \Eref{eq:csapprox} is, in general, not true. 
We discuss the limitations of this approximation in section \ref{ssec:gterm}.

Nevertheless, assuming \Eref{eq:approx} and \Eref{eq:csapprox} to hold, we could compute the integral over the source light parameters in \Eref{eq:expandintegral} to obtain
\begin{align}\label{eq:simpleintegral}
\pr(\datai|\hyperpars) = \int & d\psilens \pr(\datagal|\psilens) \pr(\teinobs|\psilens,\zsource)\prlens(\psilens|\hyperpars)\times \nonumber \\
& \prsourceeff(\zsource|\hyperpars)\pfind(\psilens,\zsource|\hyperpars)g(\psilens,\zsource),
\end{align}
where
\begin{equation}\label{eq:przsource}
\prsourceeff(\zsource|\hyperpars) = \int d\sourcelight \prsource(\zsource,\sourcelight) l(\sourcelight)
\end{equation}
is an effective probability distribution of the redshift of sources that can be strongly lensed.
One subtlety with this last definition is that $\prsourceeff(\zsource)$ is neither the redshift distribution of the background source population, which describes all sources regardless of whether they are detectable or not, nor the redshift distribution of the lensed sources. So it cannot be trivially inferred from the data without taking selection effects into account.

The factors that do not depend on the data in the integrand of \Eref{eq:simpleintegral} can be written in compact form, to indicate the effective distribution of strong lens parameters (averaged over the source light distribution):
\begin{equation}\label{eq:prsleff}
\prsleff(\psilens,\zsource) = \prlens(\psilens|\hyperpars)\prsourceeff(\zsource|\hyperpars)\pfind(\psilens,\zsource|\hyperpars)g(\psilens,\zsource).
\end{equation}
An important point is that the above probability must be properly normalised:
\begin{equation}\label{eq:norm}
\int d\psilens d\zsource \prsleff(\psilens,\zsource) = 1.
\end{equation}
The normalisation can be absorbed by the geometric cross-section factor $g$ (hence the proportionality sign in \Eref{eq:csapprox}).
With these approximations the strong lensing bias was also simplified to 
\begin{equation}\label{eq:simpleslbias}
\slbias(\psilens) \propto \int d\zsource \prsourceeff(\zsource) \pfind(\psilens,\zsource) g(\psilens,\zsource).
\end{equation}

In summary, we have approximated the expression for the lens finding probability and we have eliminated the explicit dependence of the model on the source light distribution.
The price to pay for this choice is losing sensitivity to the total number of lenses:  
had we known the number of spectroscopic detections and photometric non-detections from the SLACS search, together with the distribution of sources in redshift, surface brightness and emission line flux space, we could have used that information to further constrain the model \citep[in a similar vein to the approach of][]{ZSH24}.
We chose not to treat that aspect of the problem explicitly, for the sake of simplicity.

To compute the likelihood, one still has to evaluate the integrand of \Eref{eq:simpleintegral}.
The geometric factor $g$ can be computed by means of simulation. 
The factors $\prsourceeff(\zsource)$ and $\pfind$ can be inferred directly from the data. In \Sref{sect:model} we show a practical example of how this can be done.


\section{The model}\label{sect:model}

\subsection{Mass density profile and individual lens parameters}

Our goal was to infer the mass density profile of the population of massive galaxies, using strong lensing and stellar kinematics constraints.
We did so under the assumption of a spherical power-law density profile. 
Although the true density profile of galaxies might not be a perfect power law, deviations larger than $\sim10\%$ are inconsistent with stellar kinematics and strong lensing constraints \citep{Cap++15,Tan++24}.

A power-law profile has two degrees of freedom in the radial direction. 
We chose to parameterise it in terms of the total mass within a cylinder of radius $5$kpc, $\mfive$, and the three-dimensional density slope $\gamma$.
The choice of $\mfive$ was motivated by the fact that lensing directly constraints the projected mass, and the value of $5$~kpc is very close to the median Einstein radius of the SLACS lenses (which is $4.2$~kpc).
Therefore we expected it to be relatively well constrained by the data, without the need for major extrapolations.
Expressed in terms of $\mfive$ and $\gamma$, the three-dimensional density profile of a galaxy reads
\begin{equation}
\rho(x) = \frac{3-\gamma}{2\pi^{3/2}}\frac{\Gamma[\gamma/2]}{\Gamma[(\gamma-1)/2]}\frac{\mfive}{(5\mathrm{kpc})^{3-\gamma}}x^{-\gamma},
\end{equation}
where $x$ is the three-dimensional distance from the lens centre and $\Gamma$ is the Gamma function.

In addition to the two parameters of the power-law, we described each foreground galaxy by means of its redshift $\zlens$, its stellar mass $\mstar$, and its half-light radius $\reff$ (for simplicity also referred to as size).
We needed the redshift, first of all, to compute the lensing properties.
We also needed redshift and size (more generally, the stellar distribution) to predict the stellar velocity dispersion, to be compared with stellar kinematics measurements.
Finally, as we explain in the next section, our model for the population distribution allows for a scaling of $\mfive$ and $\gamma$ with $\mstar$ and $\reff$.
We used the stellar mass purely as a label: in our model, the value of the stellar mass of a galaxy does not directly translate into a mass density profile.
In summary, the set of individual galaxy parameters $\psilens$ is
\begin{equation}
\psilens \equiv \left\{z,m_*,\lowre,m_5,\gamma\right\},
\end{equation}
where
\begin{align}
m_* \equiv \log{\frac{\mstar}{M_\odot}}, \\
\lowre \equiv \log{\frac{\reff}{\mathrm{kpc}}}, \\
m_5 \equiv \log{\frac{\mfive}{M_\odot}}.
\end{align}

\subsection{Foreground galaxies distribution}\label{ssec:gpop}

In this section we introduce the probability distribution $\prlens$ describing the population of foreground galaxies.

We began by writing $\prlens(\psilens)$ as follows:
\begin{align}\label{eq:gpopdist}
\prlens(\psilens) = & \mathcal{M}(z,m_*)\mathcal{R}(r|m_*)\mathcal{P}(m_5,\gamma|m_*,\lowre).
\end{align}

The first factor on the right-hand side of the above equation is the distribution in redshift and stellar mass of the parent sample.
This sample is the sum of galaxies from the main and the LRG spectroscopic samples, and modelling the corresponding distribution from first principles is challenging. Nevertheless we expected, at any given redshift, the distribution to be volume-limited at the high-mass end, and to drop to zero at lower masses. Therefore we assumed the following functional form:
\begin{equation}\label{eq:mstarmodel}
\mathcal{M}(z,m_*) \propto \left(10^{m_*-\bar{m}}\right)^{\alpha+1}\exp{\left\{-10^{m_*-\bar{m}}\right\}}\frac{dV}{dz}f_t(z,m_*),
\end{equation}
where $dV/dz$ is the derivative of the comoving volume with respect to the redshift, and $f_t$ is a function that truncates the distribution at low values of $m_*$.
We chose the following form for $f_t$:
\begin{equation}\label{eq:ftrunc}
f_t(z,m_*) = \frac1\pi\arctan{\left[\frac{m_* - m_t(z)}{\sigma_t}\right]} + \frac12,
\end{equation}
where $m_t$ is a redshift-dependent truncation mass, which we model with a fifth-order polynomial function.
Because the strong lenses occupy the high-mass end of the stellar mass distribution, the exact details of the stellar mass distribution at low $\mstar$ are not critical for the accuracy of the model.

The factor $\mathcal{R}$ in \Eref{eq:gpopdist} describes the distribution in half-light radius as a function of redshift and stellar mass.
Following \citet{Son++19a}, we modelled it as 
\begin{equation}
\mathcal{R}(\lowre) = \mathcal{N}(\mu_R(m_*),\sigma_R^2),
\end{equation}
where the notation $\mathcal{N}(\mu,\sigma^2)$ indicates a Gaussian distribution with mean $\mu$ and variance $\sigma^2$.
We set the mean $\mu_R$ to scale quadratically with $\log{\mstar}$ as
\begin{equation}\label{eq:masssize}
\mu_R = \mu_{R,0} + \mu_{R,1}m_* + \mu_{R,2}m_*^2.
\end{equation}
We fixed the values of the coefficients in \Eref{eq:masssize} to those measured by \citet{H+B09}: $\mu_{R,0}=7.55$, $\mu_{R,1}=-1.84$, $\mu_{R,2}=0.11$.
Then, we set the intrinsic scatter in size to $\sigma_R=0.112$, which is the value measured by \citet{Son++19a} on a large sample of elliptical galaxies at $z=0.55$ \citep[][ did not provide an estimate of the intrinsic scatter around the mass-size relation]{H+B09}.

The last factor in \Eref{eq:gpopdist} describes the distribution in the power-law parameters, which is the main quantity of interest for our measurement.
We assumed a Gaussian functional form in both $m_5$ and $\gamma$:
\begin{equation}\label{eq:mfivegammamodel}
\pr(m_5,\gamma) = \mathcal{N}(\mu_5(m_*,\lowre),\sigma_5^2)\mathcal{N}(\mu_\gamma(m_*,\lowre),\sigma_\gamma^2).
\end{equation}
We let the average of $m_5$ and $\gamma$ to scale with stellar mass and size as
\begin{align}
\mu_5(m_*,\lowre) & = & \mu_{5,0} + \beta_5(m_* - 11.3) + \xi_5(\lowre - \mu_R(m_*)) \\
\mu_\gamma(m_*,\lowre) & = & \mu_{\gamma,0} + \beta_\gamma(m_* - 11.3) + \xi_\gamma(\lowre - \mu_R(m_*)).
\end{align}
This model is similar to that adopted by \citet{Son++13b}, but differs from it in a few aspects. 
First, here we modelled explicitly the normalisation of the total density profile, described by means of $m_5$, while \citet{Son++13b} focused only on $\gamma$.
Second, the coefficients $\xi_5$ and $\xi_\gamma$ describe a scaling with excess size, that is the ratio between the half-light radius of a galaxy and the average size of a galaxy of the same stellar mass \citep[the definition of $\xi_\gamma$ in][ is different]{Son++13b}.
Third, we did not allow for a redshift evolution, since our sample spans a much smaller redshift range than that of \citet{Son++13b}.
As we show in \Sref{sect:results}, this model is sufficiently complex to provide a good fit to the available data.

\subsection{Source galaxy distribution}\label{ssec:pzsource}

The next ingredient needed for our analysis was the redshift probability distribution of the sources that can be strongly lensed, $\prsourceeff(\zsource)$, introduced in section \ref{ssec:approx}.
According to \Eref{eq:przsource}, this quantity is the redshift distribution of the background source population, marginalised over the source light parameters, and weighted by the light-dependent component of the lens finding probability, $l$.
Our approach was to directly model $\prsourceeff(\zsource)$ empirically, so a full characterisation of $l$ was not needed.
Our model is a Gaussian distribution, with mean $\mu_{\zsource}$ and dispersion $\sigma_{\zsource}$:
\begin{equation}\label{eq:pzsource}
\prsourceeff(\zsource) = \frac{1}{\sqrt{2\pi}\sigma_{\zsource}}\exp{\left\{-\frac{(\zsource - \mu_{\zsource})^2}{2\sigma_{\zsource}^2}\right\}}.
\end{equation}

\subsection{Lensing cross-section}\label{ssec:gterm}

A key ingredient for correcting our inference for selection effects is the strong lensing cross-section, $\crosssect$.
In \Sref{ssec:approx} we split $\crosssect$ into a part that depends on the source brightness, 
and a geometric factor $g$.
This factor is the strong lensing cross-section of a source with a reference light distribution. 
In order to compute $g$, then, we needed to define exactly what a detection of a strong lens consists of, and what the light distribution of this reference source is.
The detection of SLACS lenses is both spectroscopic and photometric.
We then defined the spectroscopic detection on the basis of the total emission line flux within a circular aperture of $1.5''$ (the spectroscopic fibre), after convolution with the PSF of the SDSS, and the photometric detection based on whether there are multiple images in the HST data.
This entailed specifying both the spectrum and surface brightness distribution of the reference source.

To simplify the problem, we approximated the mass of the lenses as circular and the sources as point-like. The first assumption is justified by the fact that the strong lensing cross-section is a weak function of ellipticity; the second is a good approximation of cases in which the source size is smaller than the Einstein radius of the lens \citep[see Fig. 8 and 9 of][]{Son++23}.
Under these assumptions, we defined the emission line flux of the reference source to be $1/3$ of the total fibre flux needed for a spectroscopic detection, and the broad-band flux to be the same as that needed for a photometric detection of an image.
In other words, our reference source would be classified as strongly lensed if 1) the total emission line flux within the fibre, from all images and smeared by the PSF, is magnified by at least a factor of three with respect to the intrinsic flux; and 2) at least two images with magnification larger than one are produced.

\begin{figure}
\begin{tabular}{c}
\includegraphics[width=\columnwidth]{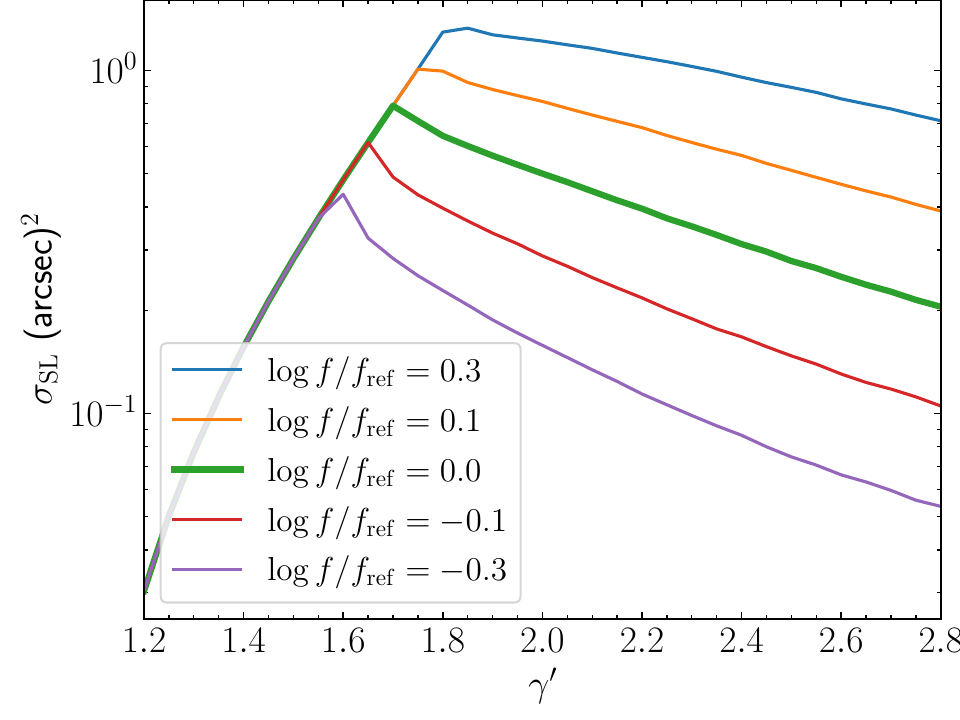} \\
\includegraphics[width=\columnwidth]{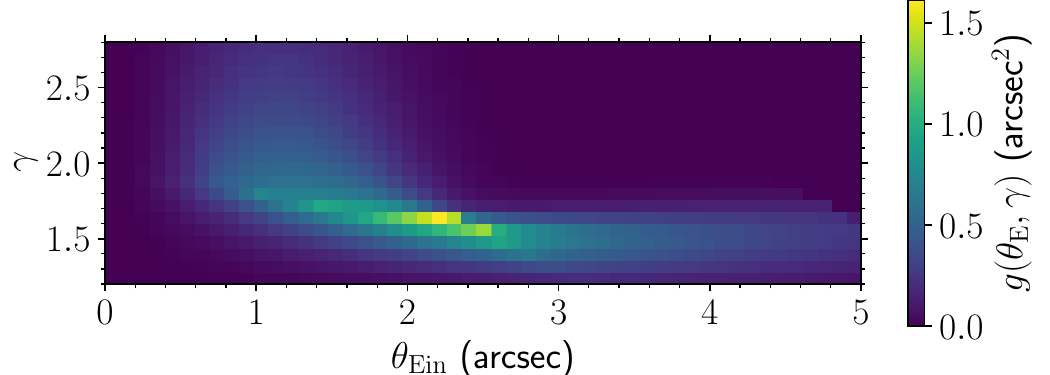}
\end{tabular}
\caption{
Strong lensing cross-section of a circular power-law lens and a point source.
A lensing event is defined as the detection of an emission line within an aperture of $1.5''$, and of at least two images in photometric data.
Top: $\crosssect$ of a lens with $\tein=1.2''$ as a function of the slope $\gamma$. The thick green line corresponds to a source with intrinsic emission line flux equal to one third of the flux required for a spectroscopic detection, and a broadband flux equal to that needed for a photometric detection. Different lines are obtained by rescaling both the emission line and broadband flux by the amount indicated in the legend.
Bottom: geometric factor of the lensing cross-section, that is the value of $\crosssect$ for the reference source (corresponding to the green line in the top panel), as a function of both $\tein$ and $\gamma$.
\label{fig:cs}
}
\end{figure}

\Fref{fig:cs} shows the strong lensing cross-section of a power-law lens with $\tein=1.2''$ with respect to the reference source (green curve), as a function of the density slope $\gamma$. We computed this quantity from \Eref{eq:csdef} by simulating lensed sources, solving the lens equation, predicting image fluxes, convolving the emission line flux by the SDSS PSF, which we took to have an FWHM of $1.5''$, and averaging over the source position.
Along with it, we also computed $\crosssect$ for sources with both emission line and broadband flux rescaled up to a factor of two with respect to the reference. This allowed us to assess the accuracy of the separability assumption of \Eref{eq:csapprox}.
For values of $\gamma > 1.8$, $\crosssect$ increases with the brightness of the source in a way that is weakly dependent on the density slope, roughly confirming our ansatz.
For smaller values, however, $\crosssect$ becomes independent of the source brightness: this is because lenses with a shallow density profile produce sets of highly magnified images, which can be detected regardless of the intrinsic brightness of the source (down to some limiting flux).
The approximation of \Eref{eq:csapprox} therefore breaks down for very bright sources or for lenses with a very shallow density profile.
More generally, the accuracy of our approximation is higher the more similar our reference source is to the typical strongly lensed source of the SLACS sample.
Since the choice of the reference source is arbitrary, we also carried out the analysis with a different definition of it, by setting its emission line flux to half of that necessary for a spectroscopic detection (instead of a third). The results did not change appreciably.

The bottom panel of \Fref{fig:cs} shows the geometric factor of the lensing cross-section, as a function of $\tein$ and $\gamma$, that is $\crosssect$ of our reference source.
For $\gamma > 2$, the lensing cross-section peaks at values of the Einstein radius close to the fibre size, then drops to zero for larger values of $\tein$. This is a distinct signature of the spectroscopic detection requirement:
for lenses with a steep density profile and large Einstein radius, any images that form within the spectroscopic fibre are not sufficiently magnified to lead to a detection \citep[see also][]{Arn++12}.

\subsection{Lens finding probability}

Finally, we needed to prescribe a model for the probability of finding a detectable lens.
As we explained in section \ref{ssec:slacssel}, the main quantity that determined the discovery of a SLACS lens was the estimated lensing cross-section, which is the square of \Eref{eq:teinest}.
Therefore, we asserted that
\begin{equation}
\pfind(\psilens,\psisource) = \pfind\left(\teinest\right),
\end{equation}
which is a special case of \Eref{eq:approx}, since $\teinest$ depends only on lens properties and on the source redshift (see \Eref{eq:teinest}).
Our choice for the functional form of $\pfind$ is the following sigmoid, described by two parameters:
\begin{equation}\label{eq:ffind}
\pfind(\teinest) = \frac{1}{1 + \exp{\left[-a(\teinest - \theta_0)\right]}}.
\end{equation}
According to our model, the lens finding probability is zero for small values of the estimated Einstein radius, and reaches one for values of $\teinest$ larger than $\theta_0$.

Predicting $\teinest$ for the parent population of lenses requires accounting for the effects of observational noise on the velocity dispersion measurements. We adopted a $6.25\%$ uncertainty on all $\sigmaap$ measurements, which is the median relative uncertainty on $\sigmaap$ of the SLACS lenses.
By comparison, \citet{Bol++08} quoted a value of $7\%$ as representative of the uncertainty on $\sigmaap$ measurements of the SLACS lenses.
\citet{Bir++20} argued, on the basis of a mismatch between the inferred properties of the SLACS and a different set of lenses, that the velocity dispersion measurements of SLACS lenses have been underestimated. They allowed for a systematic uncertainty, which they constrained to be approximately $6\%$, to be added in quadrature to the SLACS values. We repeated the analysis with these added uncertainties, but found that the goodness-of-fit worsened. Therefore we stuck with the uncertainties provided by \citet{Aug++09} for our analysis.

\subsection{Fundamental plane constraint}\label{ssec:fp}

Our goal was to infer the properties of the parent population of galaxies, and we could use information on the non-lens galaxies for this purpose.
The model that we introduced so far makes use of the stellar mass-redshift distribution of the parent population of the lenses, and of the mass-size relation of early-type galaxies.
A natural extension would have been to also consider the velocity dispersion distribution.
Our model makes predictions for the velocity dispersion of a galaxy, by means of Jeans modelling. 
Therefore, it is in principle possible to fit the model directly to the observed velocity dispersion measurements of the entire parent population.
We chose not to, for the following reasons.
First, with a sample of $\sim4\times10^5$ objects, the constraints from the dynamical model of the parent population would dominate over the lensing data. Second, the simplified assumptions on which our dynamical analysis is based would end up completely dominating the error budget: we would need to adopt a much more complex model of the stellar orbits and of the measurement systematics to properly fit such a large dataset. Third, the inference is sensitive to the input values of the stellar mass, size and velocity dispersion of the galaxies, and to their uncertainties. Since the stellar mass and half-light radius of the lenses and non-lenses have been measured with different procedures, separate dynamical analyses of the two samples can easily produce inconsistent posterior probabilities, which therefore cannot be combined.

In light of these difficulties, our compromise solution was to adopt a prior on the velocity dispersion distribution of the parent population that matches the observations, but allowing for some degree of flexibility.
In particular, we set a prior on the coefficients of the stellar mass fundamental plane, the observed correlation between stellar mass, half-light radius and central velocity dispersion of early-type galaxies \citep[see e.g.][]{Ber++20}.
We defined the stellar mass fundamental plane as follows:
\begin{equation}\label{eq:fpmodel}
\log(\sigmaap) \sim \mathcal{N}(\mu_{\mathrm{FP},0} + \beta_{\mathrm{FP}}(m_* - 11.3) + \xi_{\mathrm{FP}}(\lowre - \mu_{R}(m_*)), \sigmafp^2).
\end{equation}
We could not fit directly \Eref{eq:fpmodel} to the parent sample galaxies, because we lacked measurements of their half-light radius (or rather, we lacked measurements that are consistent with those of the SLACS lenses).
So, we fitted the stellar mass-velocity dispersion relation instead, to constrain the first two coefficients in \Eref{eq:fpmodel}.
We assumed a quadratic dependence of $\log{\sigmaap}$ on $m_*$ for the fit.
We obtained $\mu_{\mathrm{FP},0} = 2.342$ and $\beta_{\mathrm{FP}} = 0.258$ for galaxies with stellar mass close to the pivot value of $m_*=11.3$, and adopted a Gaussian prior centred on these values. The statistical errors on these coefficients are very small, but we allowed for a $0.03$ dispersion to accommodate possible inconsistencies between the measurement on the SLACS sample and the parent population.
Finally, we put a prior on the intrinsic scatter parameter $\sigmafp$. This is an important quantity that defines the tightness of the fundamental plane relation. We defined our prior on $\sigmafp$ in a way to accommodate both measurements from the literature and from the SLACS sample.
\citet{Son20} measured $\sigmafp=0.047\pm0.02$ on a sample of early-type galaxies at $z=0.2$, and, when fitting \Eref{eq:fpmodel} to the SLACS data used in this paper, we obtained $\sigmafp=0.039\pm0.008$.
On the basis of these results, we chose the following prior:
\begin{align}\label{eq:fpprior}
\pr(\mu_{\mathrm{FP},0}) = \mathcal{N}(2.342,0.030^2) \nonumber \\
\pr(\beta_{\mathrm{FP}}) = \mathcal{N}(0.258,0.030^2) \\
\pr(\sigmafp) = \mathcal{N}(0.047, 0.008^2) \nonumber.
\end{align}
In practice, for a given value of the model parameters $\hyperpars$ we fitted \Eref{eq:fpmodel} to a large mock population of galaxies, then weighted the corresponding posterior probability by \Eref{eq:fpprior}. 
Since we were mostly interested in obtaining an accurate model for the high-mass end of the galaxy distribution, where the bulk of the SLACS lenses lie, we restricted this fit to the stellar mass range $m_* > 11.0$.

We found that the prior of \Eref{eq:fpprior} had a significant impact on our inference. 
Therefore, in \Sref{sect:results}, we show results obtained when both including or excluding it.

\subsection{Inference procedure}

Our fiducial model is described by the following set of free parameters:
\begin{align}
\hyperpars \equiv \{ & \bar{m},\alpha,m_t^{(0)},\ldots,m_t^{(5)},\sigma_t,\mu_{5,0},\beta_5.\xi_5, \mu_{\gamma,0}, \beta_{\gamma},\xi_{\gamma},\sigma_\gamma,\mu_{\zsource},\nonumber \\
& \sigma_{\zsource},\theta_0,a\},
\end{align}
where $m_t^{(0)},\ldots,m_t^{(5)}$ are the coefficients of the polynomial function describing the truncation mass $m_t$ in \Eref{eq:ftrunc}.
We first constrained the stellar mass function parameters, by fitting the model of \Eref{eq:mstarmodel} to the stellar mass measurements of the parent population. The inferred values are reported in \Tref{tab:mz}. In virtue of the very large sample size, formal uncertainties are very small. Therefore we fixed these parameters to their best-fit value and proceeded to infer the remaining parameters by fitting the model to the SLACS lenses.

\begin{table}
\caption{Parent sample. Stellar mass and redshift distribution parameters.}
\label{tab:mz}
\begin{tabular}{cc}
Parameter & Value \\
\hline
\hline
$\bar{m}$ & $11.06$\\
$\alpha$ & $-1.207$\\
$m_t^{(0)}$ & $9.388$\\
$m_t^{(1)}$ & $7.855$\\
$m_t^{(2)}$ & $48.34$\\
$m_t^{(3)}$ & $-312.5$\\
$m_t^{(4)}$ & $535.7$\\
$m_t^{(5)}$ & $-274.2$\\
$\sigma_t$ & $0.0007$\\

\end{tabular}
\tablefoot{
Inferred parameter values of the model describing the distribution in redshift and stellar mass of the parent sample. The parameters are defined in \Eref{eq:mstarmodel} and \Eref{eq:ftrunc}.
}
\end{table}

The data vector of each SLACS lenses consists of the lens redshift, source redshift, lens stellar mass, half-light radius, stellar velocity dispersion, and Einstein radius:
\begin{equation}
\datai \equiv \{z_{\mathrm{g}},m_{*}^{(\mathrm{obs})},\lowre,\sigma_{\mathrm{ap}}^{(\mathrm{obs})},\teinobs,\zsource\}.
\end{equation}
Stellar mass and velocity dispersion have associated uncertainties, on the order of $25\%$ and $6\%$ respectively. Redshift and half-light radius are instead measured very precisely, and we neglected their uncertainties.
Einstein radii are measured with a precision of a few percent \citep{Eth++22}. To further simplify the inference procedure, we neglected their uncertainties as well. We justified this approximation after the inference, by noting that our measurement on the average mass within $5$~kpc has a larger relative uncertainty, which is therefore dominated by other aspects of the data (most likely the velocity dispersion measurements).

We assumed a Gaussian form for the likelihood of the stellar mass and velocity dispersion measurements.
We calculated the model stellar velocity dispersion, $\sigmaap$, using the spherical Jeans equation under the assumption of isotropic orbits. We discuss the possible implications of these assumptions in \Sref{sect:discuss}.

From \Eref{eq:simpleintegral} and \Eref{eq:prsleff} we could write the likelihood term for a SLACS lens as 
\begin{align}\label{eq:lenslike}
\pr(\datai|\hyperpars) = \int & d m_* d\gamma dm_5 \pr(\mlowstarobs|m_*)\pr(\sigmaapobs|\sigmaap(m_5,\gamma)) \times \nonumber \\
& \pr(\teinobs|\tein(m_5,\gamma))\prsleff(m_*,m_5,\gamma,\zsource|\hyperpars).
\end{align}
Under the assumption of negligible uncertainty on the Einstein radius, the likelihood term in $\teinobs$ became a Dirac delta function, which we could integrate over after a variable change:
\begin{equation}\label{eq:varchange}
(\gamma,m_5) \rightarrow (\gamma,\tein(\gamma,m_5)).
\end{equation}
The result is 
\begin{align}\label{eq:lenslike2}
\pr(\datai|\hyperpars) = \int & d m_* d\gamma \pr(\mlowstarobs|m_*)\pr(\sigmaapobs|\sigmaap(\mfiveobs(\gamma),\gamma)) \times \nonumber \\
& \prsleff(m_*,\mfiveobs(\gamma),\gamma,\zsource|\hyperpars) \left|\frac{d m_5}{d \tein}\right|_{\tein=\teinobs},
\end{align}
where $\mfiveobs(\gamma)$ is the value of $m_5$ that reproduces the observed Einstein radius, for a given $\gamma$.

We assumed a uniform prior in all model hyper-parameters, multiplied by the fundamental plane prior of \Eref{eq:fpprior}.
We sampled the posterior probability distribution with a Markov Chain Monte Carlo (MCMC).
We carried out the integrals of \Eref{eq:lenslike}, 
as well as the integral of \Eref{eq:norm} to normalise $\prsleff$, via Monte Carlo integration.

\subsection{Lens-only inference}\label{ssec:slacsonly}

One of the goals of this work was to quantify the impact of selection effects on a strong lensing measurement of the mass-structure of galaxies.
To do this, we also carried out the inference while ignoring all lensing selection terms. 
We did this by taking the prior factor $\prsleff$ in \Eref{eq:lenslike2} to simply be the product between the distribution in $(m_5,\gamma)$ of \Eref{eq:mfivegammamodel} and a Gaussian distribution in $m_*$, meant to describe solely the SLACS lenses.


\section{Results}\label{sect:results}

\subsection{Individual lens constraints}

\begin{figure}
\includegraphics[width=\columnwidth]{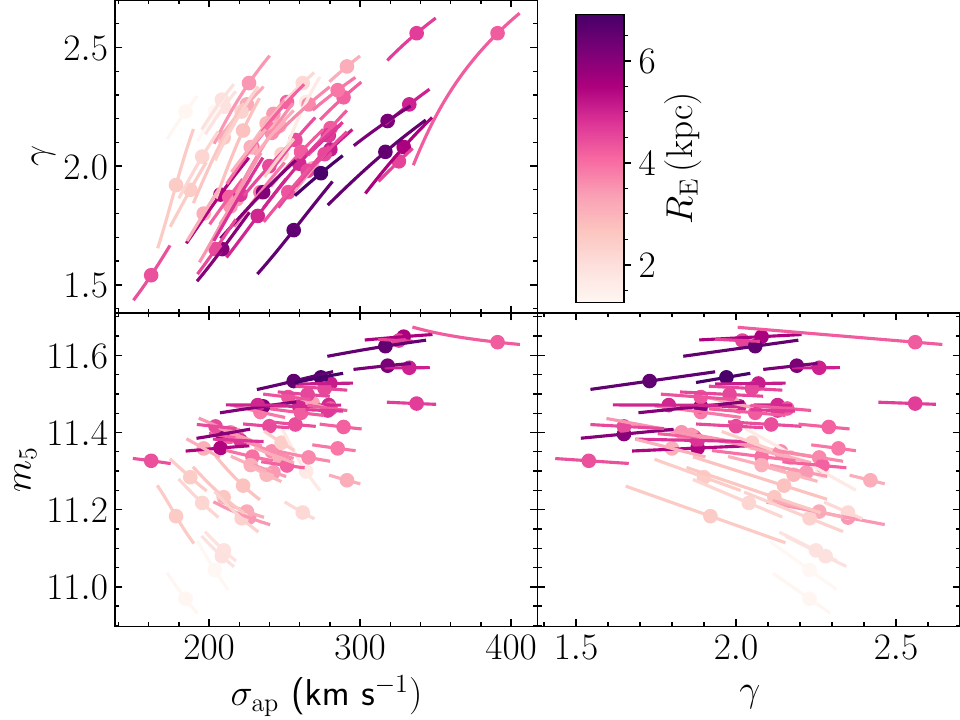}
\caption{
Constraints on $m_5$ and $\gamma$, and predicted values of the velocity dispersion, for individual lenses.
Each line covers the 68\% credible region of the posterior probability.
Lines are colour-coded by the physical size of the Einstein radius.
Dots mark the maximum posterior probability values.
\label{fig:individ}
}
\end{figure}

As a first thing we examined the constraining power of each individual lens.
A power-law model has two free parameters: $m_5$ and $\gamma$.
Each lens provided two constraints: the Einstein radius, which we assumed to be measured exactly, and the stellar velocity dispersion within the spectroscopic aperture, $\sigmaap$.
Therefore, the range of parameter values allowed by the data is a line in $m_5-\gamma$ space. 
These constraints are shown in \Fref{fig:individ}, along with the values of the velocity dispersion corresponding to the model. Each line spans the 68\% credible region of the posterior probability, assuming a flat prior on both $m_5$ and $\gamma$.
The mass enclosed within $5$~kpc is generally very well constrained: the median uncertainty on $m_5$ is as small as $0.013$. 
The uncertainty is largest for the lenses with Einstein radius that differs the most from $5$~kpc: for these lenses, obtaining $m_5$ requires an extrapolation of the power-law profile, which introduces a correlation between $\gamma$ and $m_5$.
In all cases, $\gamma$ is strongly correlated with the velocity dispersion: given the Einstein radius of a lens, the steeper the density profile the larger the value of $\sigmaap$.
Together with the velocity dispersion selection criterion that we described in section \ref{ssec:slacssel}, this has important implications for the selection function of the SLACS sample.

\subsection{Parameter inference}

\begin{figure*}
\includegraphics[width=\textwidth]{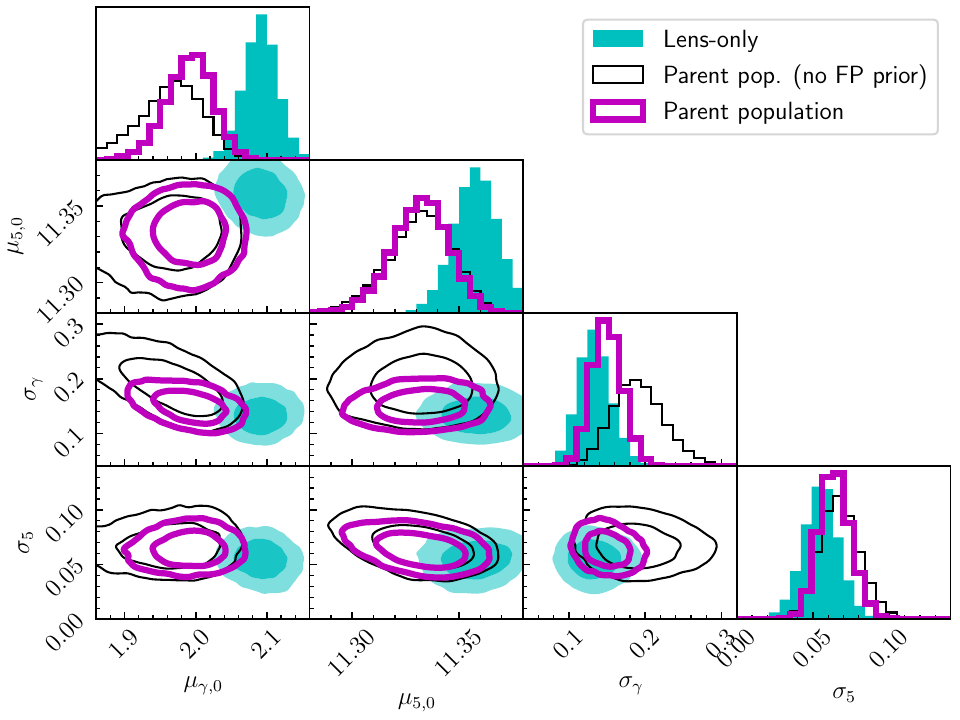}
\caption{
Posterior probability distribution of the model.
Parameters shown: average $\log{\mfive}$ and $\gamma$ at fixed stellar mass and half-light radius, and intrinsic scatter around the average.
Filled contours: model describing SLACS lenses, without correction for selection effects (see section \ref{ssec:slacsonly}).
Thick solid curves: fiducial model.
Thin solid curves: model with no fundamental plane prior of \Eref{eq:fpprior}.
Contour levels correspond to 68\% and 95\% enclosed probability.
\label{fig:mupars}
}
\end{figure*}

\begin{figure*}
\includegraphics[width=\textwidth]{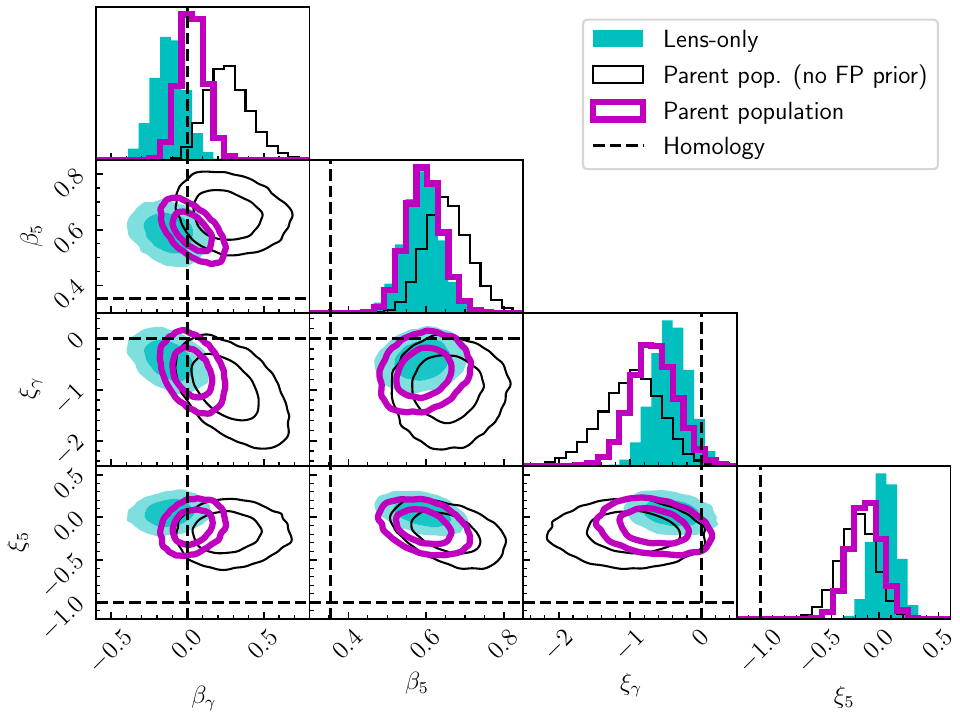}
\caption{
Same as \Fref{fig:mupars}, but showing parameters describing the scaling of $m_5$ and $\gamma$ with stellar mass and size.
Dashed lines indicate values of the parameters corresponding to the case in which galaxies are homologous systems with an average density slope of $\gamma=2.00$.
\label{fig:deppars}
}
\end{figure*}

\begin{figure}
\includegraphics[width=\columnwidth]{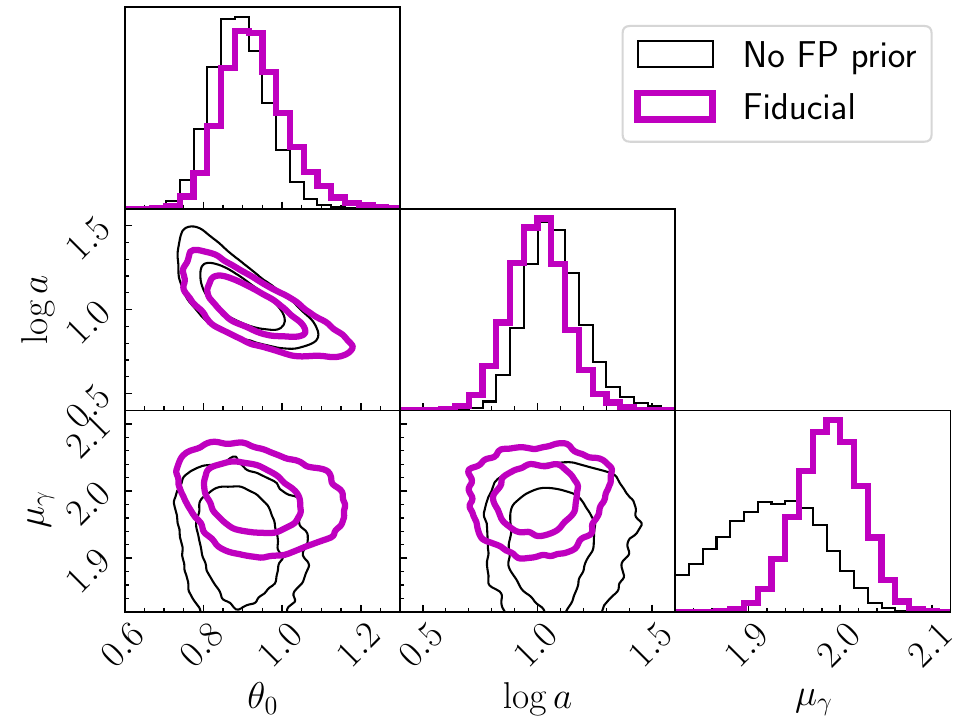}
\caption{
Same as \Fref{fig:mupars}, but showing parameters describing the lens finding probability, defined in \Eref{eq:ffind}, and the average density slope parameter.
\label{fig:findpars}
}
\end{figure}

In \Tref{tab:results} we report the median and $68\%$ credible region of the marginal posterior probability of each model parameter.
We list the values obtained with and without the fundamental plane prior of \Eref{eq:fpprior}, as well as the values obtained when fitting the SLACS lenses while not accounting for selection effects.
In \Fref{fig:mupars}, \Fref{fig:deppars} and \Fref{fig:findpars} we show projections of the posterior probability on selected pairs of model parameters. 
Given the large number of free parameters, we avoid showing all possible pairwise correlations. The MCMC samples that we used to make these figures is available online for further analysis.

\begin{table*}
\caption{Results.}
\label{tab:results}
\begin{tabular}{ccccl}
Parameter & Fiducial model & No FP prior & Lens-only & Parameter description \\
\hline
\hline
$\mu_{5,0}$ & $11.332 \pm 0.013$ & $11.338 \pm 0.015$ & $11.358 \pm 0.010$ & Mean $\log{\mfive}$ at $\log{\mstar}=11.3$ and average size \\
$\beta_{5}$ & $0.59 \pm 0.05$ & $0.65 \pm 0.06$ & $0.59 \pm 0.05$ & Dependence of the mean $\log{\mfive}$ on $M_*$ \\
$\xi_{5}$ & $-0.11 \pm 0.13$ & $-0.22 \pm 0.16$ & $0.05 \pm 0.10$ & Dependence of the mean $\log{\mfive}$ on excess size \\
$\sigma_{5}$ & $0.064 \pm 0.010$ & $0.062 \pm 0.013$ & $0.055 \pm 0.011$ & Scatter in $\log{\mfive}$ around the mean \\
$\mu_{\gamma}$ & $1.99 \pm 0.03$ & $1.97 \pm 0.04$ & $2.091 \pm 0.024$ & Mean $\gamma$ at $\log{\mstar}=11.3$ and average size \\
$\beta_{\gamma}$ & $0.03 \pm 0.08$ & $0.31 \pm 0.15$ & $-0.12 \pm 0.10$ & Dependence of the mean $\gamma$ on $M_*$ \\
$\xi_{\gamma}$ & $-0.67 \pm 0.31$ & $-1.05 \pm 0.43$ & $-0.42 \pm 0.26$ & Dependence of the mean $\gamma$ on excess size \\
$\sigma_{\gamma}$ & $0.149 \pm 0.020$ & $0.19 \pm 0.04$ & $0.133 \pm 0.022$ & Scatter in $\gamma$ around the mean \\
$\mu_{z_{\mathrm{s}}}$ & $0.48 \pm 0.04$ & $0.518 \pm 0.018$ & $\ldots$ & Mean of the effective source redshift distribution \\
$\sigma_{z_{\mathrm{s}}}$ & $0.215 \pm 0.021$ & $0.212 \pm 0.021$ & $\ldots$ & Dispersion of the effective source redshift distribution \\
$\theta_0$ & $0.93 \pm 0.08$ & $0.80 \pm 0.06$ & $\ldots$ & Lens finding probability parameter \\
$\log{a}$ & $1.00 \pm 0.12$ & $1.42 \pm 0.26$ & $\ldots$ & Lens finding probability parameter \\

\end{tabular}
\tablefoot{
Inference.
}
\end{table*}

\Fref{fig:mupars} focuses on the mean $m_5$ and $\gamma$ at fixed stellar mass and size, and scatter around the average.
According to the lens-only inference (cyan filled contours), SLACS lenses of $m_*=11.3$ and average size for their mass have an average density slope of $\mu_{\gamma,0} = 2.091\pm0.024$. 
This value is in good agreement with the measurement of \citet{Aug++10} on the same dataset.
Once corrected for selection effects, the same quantity for the parent sample galaxies (magenta solid curves) is $\mu_{\gamma,0} = 1.99\pm0.03$.
The average projected mass within $5$~kpc, at the pivot stellar mass and average size, is inferred to be $\mu_{5,0}=11.358\pm0.010$ in the lens-only analysis and $\mu_{5,0}=11.332\pm0.013$ after the selection effects correction.
Based on these results, we can say that SLACS lenses are on average slightly more massive than parent sample galaxies (they have a larger $m_5$ at fixed stellar mass), and have an overall steeper density profile. The larger mass can be easily explained by the increase in lensing cross-section with $m_5$. The bias on $\gamma$ is less intuitive, because, as the bottom panel of \Fref{fig:cs} shows, the lensing cross-section is generally larger for values of $\gamma < 2$: this alone would lead to the lenses having a shallower profile than the non-lenses. Instead, as we show in section \ref{ssec:insight}, the main reason for the bias in $\gamma$ lies in the lens finding probability term, $\pfind$.

When comparing the inference based on the fiducial model (magenta curves) with that without the fundamental plane prior (black curves), we can see that the strong lensing bias (the difference with respect to the SLACS-only posterior probability) is larger in the latter case.
To understand this fact, it is useful to consider the intrinsic scatter parameters.
We can see that $\sigma_\gamma$ and $\sigma_5$ are anti-correlated with the respective average parameters, $\mu_{\gamma,0}$ and $\mu_{5,0}$.
This means that the strong lensing data can be reproduced both with an underlying population that has a relatively large bias and intrinsic scatter, or with a smaller bias and scatter. But a larger scatter in $m_5$ and $\gamma$ translates into a larger scatter around the fundamental plane. Therefore the solution with the larger bias is disfavoured by the prior of \Eref{eq:fpprior}.

\Fref{fig:deppars} shows the posterior probability distribution in the parameters describing how $m_5$ and $\gamma$ scale with stellar mass and size.
In order to gain intuition on how to interpret these measurements, it is useful to first predict the values of these parameters in an ideal case.
For instance, we can make the assumption that early-type galaxies are homologous systems, which is to say that galaxies with different values of the stellar mass and half-light radius are scaled-up versions of one another.
Although strict homology is already ruled out by dynamical observations, such as the tilt of the fundamental plane \citep[see e.g.][]{BCD02}, we can still use it as a reference limiting case.
Under the homology assumption, the average density slope $\gamma$ is a constant, the total projected mass within the half-light radius is also a constant, and, at fixed half-light radius, the total mass within any aperture scales linearly with stellar mass.
The values of the parameters of our model under the homology assumption are $\beta_5 = 0.35$, $\xi_5=-1.00$, $\beta_\gamma=0$ and $\xi_\gamma=0$.

Our measured values are inconsistent with the homology assumption.
In particular, $\beta_5 = 0.59\pm0.05$, which means that the mass enclosed within $5$~kpc grows more rapidly than it would in the case of perfect homology. It could be explained with the dark matter mass within the half-light radius growing more than linearly with stellar mass, or also with the ratio between the true stellar mass and our estimate $m_*$ being an increasing function of $m_*$.
This is not a new result, but was previously noted by \citet{Aug++10} in their analysis of SLACS lenses.
Our work demonstrates that it holds also when accounting for selection effects.

The dependence of $m_5$ on excess size is also very different from that of homologous systems: $\xi_5 = -0.11\pm0.13$, instead of $-1$. 
This means that $m_5$ responds relatively mildly to a change in half-light radius. One possible explanation is that the dark matter distribution is roughly independent of galaxy size. Weak lensing observations have shown that halo mass is a weak function of size, at fixed stellar mass \citep{Son++18,Son++19a,Son++22}, so they support this hypothesis \citep[though see][ for a counter-example]{Cha++17}.
Finally, we put an upper limit on the correlation between the density slope $\gamma$ and stellar mass, $\beta_\gamma=0.03\pm0.08$, and found a negative correlation with galaxy size at fixed stellar mass: $\xi_\gamma = -0.67\pm0.31$. This is consistent with the measurements of \citet{Son++13b}.

Among the model parameters are those describing the effective source redshift distribution, $\prsourceeff$, introduced in section \ref{ssec:pzsource}.
The mean and dispersion of the Gaussian of \Eref{eq:przsource} are constrained to be $\mu_{\zsource} = 0.48\pm0.04$ and $\sigma_{\zsource} = 0.22\pm0.02$, and the two parameters are anti-correlated. By comparison, the mean and standard deviation of the redshift of the SLACS sources is $0.61$ and $0.18$, respectively. 
The fact that we inferred a broader distribution, centred at a smaller redshift, was expected: as we explained in section \ref{ssec:approx}, $\prsourceeff$ describes the redshift distribution of unlensed sources that can be detected. 
Because the lensing cross-section increases for higher-redshift sources, the distribution of SLACS sources is shifted to higher redshifts.
We found no signs of correlation between $\mu_{\zsource}$ or $\sigma_{\zsource}$ and any of the other parameters of the model. From this we concluded that the exact choice of parameterisation of $\prsourceeff$ is not a critical element in our model.

Finally, we constrained the lens finding probability parameters, defined in \Eref{eq:ffind}. \Fref{fig:findpars} shows the posterior probability in the cutoff radius $\theta_0$ and $\log{a}$, defined in \Eref{eq:ffind}, as well as in the average density slope $\mu_{\gamma,0}$.
The cutoff on $\teinest$ is $\theta_0=0.93\pm0.08$, meaning that lenses with an estimated Einstein radius smaller than this value had a small probability of being selected for photometric follow-up in the SLACS survey. 
The parameters $\theta_0$ and $a$ are anti-correlated. This indicates that the data cannot constrain the detailed shape of the lens finding probability function, but can allow either a $\pfind$ with a sharp cutoff at a smaller $\teinest$, or one with a more progressive increase centred at a larger value of $\teinest$.
Neither $\theta_0$ nor $\log{a}$ are strongly correlated with the average density slope parameter $\mu_{\gamma,0}$, as can be seen in \Fref{fig:findpars}, or with any other parameter of the model.

\subsection{Goodness of fit}

We assessed the goodness of fit of the model with posterior predictive tests, as follows.
We sampled sets of values of the model parameters $\hyperpars$ from the posterior probability distribution. For each set of values, we randomly drew a set of $59$ lenses (the same as the SLACS sample) from the strong lens distribution $\prsleff$. For each lens, we predicted the set of observables that we used in the analysis, adding observational noise to the velocity dispersion. We adopted a Gaussian noise with a $6.25\%$ relative uncertainty for this purpose.
We then defined an ensemble of test quantities $\{T_i\}$ based on observations, and quantified the probability of drawing more extreme values of each test quantity from a posterior predicted sample.
Very small ($\pr < 5\%$) or very large ($\pr > 95\%$) probabilities would mean that it is unlikely that the observed sample has been drawn from the model.
The choice of test quantities is not unique, but must be made on the basis of what aspects of the data that we want the model to successfully reproduce.
We chose them to be the mean, standard deviation, $10\%$- and $90\%$-ile of the distribution in observed Einstein radius, velocity dispersion and density slope $\gamma$.

\begin{figure*}
\includegraphics[width=\textwidth]{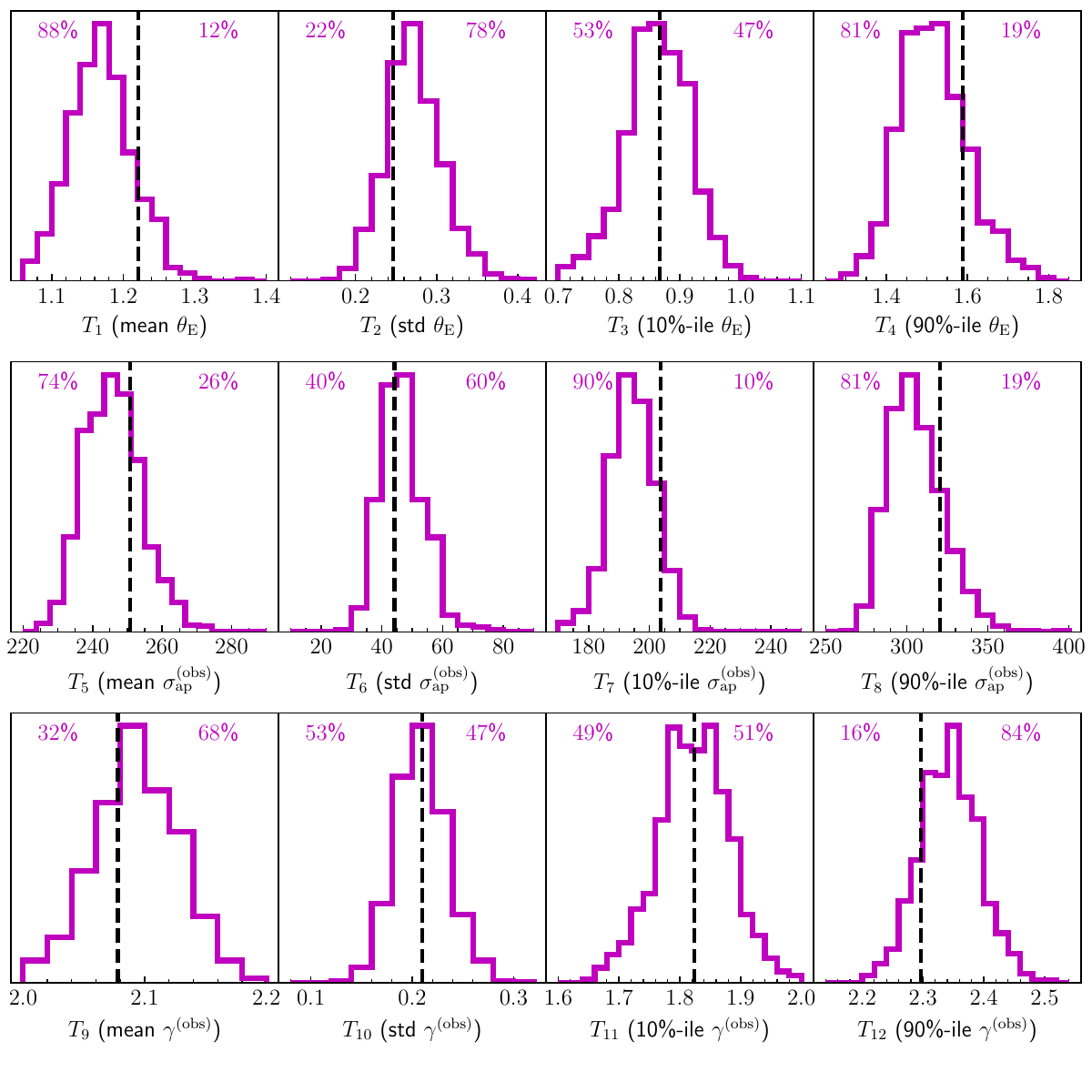}
\caption{
Posterior predictive tests of goodness of fit.
Each panel shows the posterior predicted (histogram) and observed value (dashed line) of the test quantity indicated at the bottom. Units of Einstein radius are in arcsec and of the velocity dispersion are in km~s$^{-1}$.
\label{fig:pp}
}
\end{figure*}

\Fref{fig:pp} shows the posterior predicted distributions of these twelve test quantities.
The probability of producing samples of lenses with more extreme values of each test quantity is shown in the corresponding panel of \Fref{fig:pp}. None of the observed test quantities correspond to very small or very large percentiles of the posterior predicted distribution. Therefore our model is able to reproduce these data.

We also checked whether the model can recover the velocity dispersion distribution of the parent sample. 
The bottom panel of \Fref{fig:trends} shows the posterior predicted average $\sigmaap$ (magenta band) as a function of stellar mass, 
and our quadratic fit to the $m_*-\sigmaap$ distribution of the parent sample (green curve).
The model matches the data over a broad range in stellar mass, but departs from it at the high-mass end.
This is not a goodness-of-fit issue: as we explained in section~\ref{ssec:fp}, we only used observations around the pivot stellar mass $m_*=11.3$ to constrain the model.
Nevertheless, it could mean that the model does not extrapolate well at large stellar masses, where it tends to overpredict the stellar velocity dispersion.
However, the source of the discrepancy could more simply be an inconsistency between the way stellar mass is defined for the SLACS lenses and the parent sample galaxies. Although we rescaled the $m_*$ measurements of the parent sample galaxies so that the stellar masses of the SLACS lenses would be consistent on average, we cannot exclude a mass-dependent offset, for instance as a result of different choices in the photometric measurement or stellar population synthesis phase.
It could also be that a quadratic model is not sufficiently flexible to describe the high-mass end of the $m_*-\sigmaap$ relation.
Our main goal is to understand the properties of galaxies at $m_*\approx11.3$, where the bulk of the SLACS lenses lie. Therefore we did not investigate this issue further.

\subsection{Insight from the model}\label{ssec:insight}

\begin{figure}
\includegraphics[width=\columnwidth]{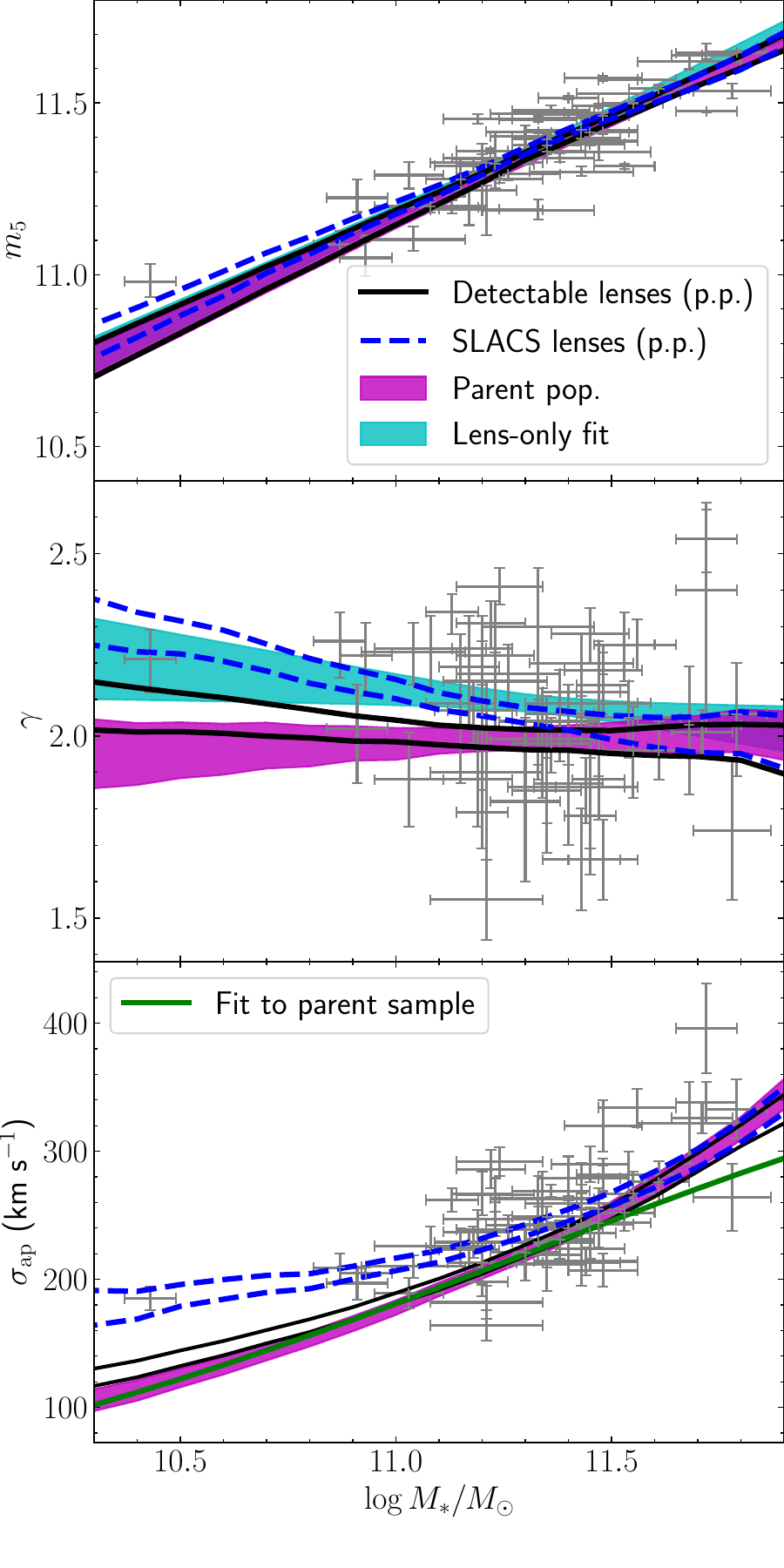}
\caption{
Posterior predicted trends with stellar mass, for galaxies of average size.
Top: average projected mass enclosed within $5$~kpc.
Middle: average density slope.
Bottom: average stellar velocity dispersion.
Each band marks the 68\% credible region.
The magenta band shows the prediction for the parent population.
The cyan band shows the prediction from the lens-only analysis, without accounting for selection effects.
Dashed blue lines show the prediction for the population of SLACS lenses.
Solid black lines show the prediction for the population of detectable lenses (obtained by removing the lens finding probability $\pfind$ from the posterior prediction).
Error bars are the values measured for the SLACS lenses, assuming a flat prior on both $m_5$ and $\gamma$. Uncertainties on the stellar mass are uncorrelated with those on $m_5$, $\gamma$ and $\sigmaap$.
In the bottom panel, the green line shows our quadratic fit to the $m_*-\sigmaap$ relation of the parent sample.
\label{fig:trends}
}
\end{figure}

\Fref{fig:trends} shows the inferred trend of $m_5$, $\gamma$ and velocity dispersion with stellar mass, as predicted from the posterior probability of the model parameters, along with the data.
In the middle panel we can see a significant difference between the average $\gamma$ of the parent population (magenta band), and that inferred from the lens-only analysis (cyan band).
According to these results, the average density slope of the SLACS lenses appears to be $\sim0.1$ larger than the corresponding value of their parent population, at the pivot stellar mass $m_*=11.3$.
This raises the question of whether this bias is intrinsic to strong lenses in general, or is a peculiarity of SLACS.

One aspect that makes the SLACS sample unique is the pre-selection of targets for photometric follow-up based on the estimated Einstein radius, which we described empirically with the lens finding probability term $\pfind$.
We can assess the impact of this choice with posterior prediction.
For a given galaxy population drawn from the posterior probability, we considered the population of detectable lenses: these are strong lens systems that are detected spectroscopically and that would be detected photometrically if followed-up with HST.
We obtained such a sample by setting $\psel = \pdet$, dropping $\pfind$ in \Eref{eq:two}.
We measured the average $m_5$ and $\gamma$ of detectable lenses in bins of stellar mass. The resulting $68\%$ credible regions are shown in \Fref{fig:trends}, enclosed within the black lines.
At stellar masses around $m_* \sim 11.3$, where the bulk of the SLACS data lies, the posterior predicted average $\gamma$ of detectable lenses is indistinguishable from the value inferred for the lens population (magenta band).
This means that, according to our model, the strong lensing bias on the density slope is entirely attributable to the $\teinest$-based target prioritisation criterion.
At fixed lens and source redshift, selecting lenses with larger $\teinest$ means selecting preferentially lens galaxies with larger velocity dispersion. Since velocity dispersion correlates positively with $\gamma$, this results in lenses with steeper density profiles on average.

The above argument explains part of the difference in the average density slope that we inferred with or without taking selection effects into account, but not its entirety.
From our model, we predicted the average $\gamma$ of SLACS lenses, this time using the full expression for $\psel$.
The corresponding 68\% confidence region is show by the dashed blue lines in \Fref{fig:trends}.
This is indeed larger than the corresponding value for the lens population, but is also smaller than the value that we inferred from the lens-only analysis (cyan band in \Fref{fig:trends}). The difference in $\gamma$ is about $0.04$ at the pivot stellar mass $m_*=11.3$.
This result might seem counterintuitive, especially given that the posterior predicted tests showed no sign of inconsistency with the observed values of $\gamma$ (bottom panels of \Fref{fig:pp}).

\begin{figure}
\includegraphics[width=\columnwidth]{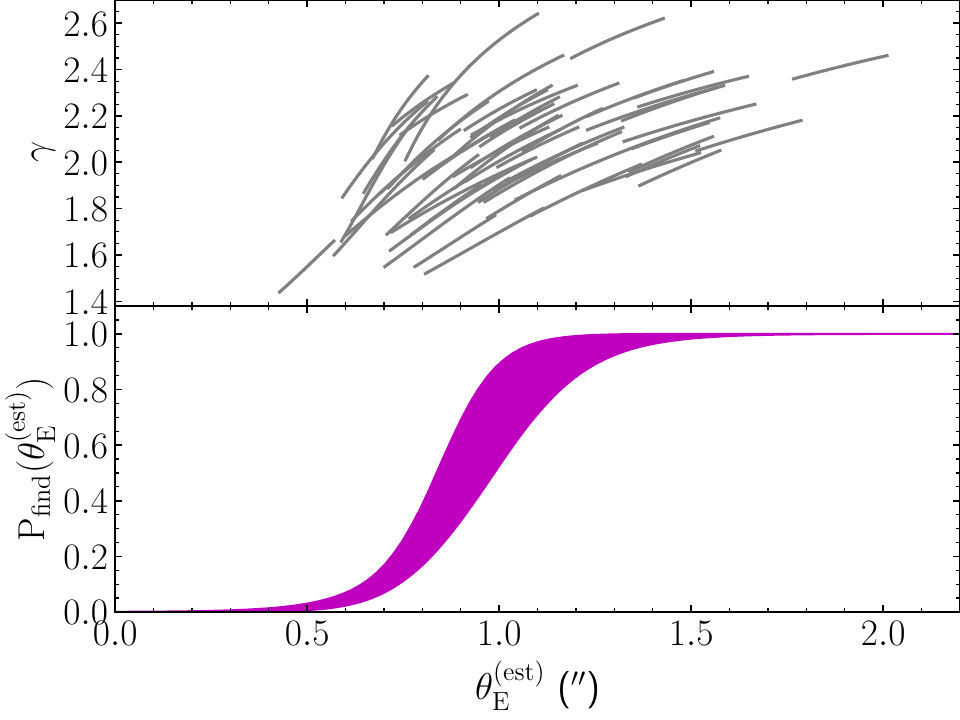}
\caption{
Estimated Einstein radius, density slope and lens finding probability.
Top: 68\% probable values of $\teinest$ and density slope $\gamma$ of the SLACS lenses, where $\teinest$ is defined in \Eref{eq:one}.
Bottom: 68\% credible region of the lens finding probability, defined in \Eref{eq:ffind}.
The correlation between $\teinest$ and $\gamma$ makes so that lenses with overestimated $\gamma$ are more likely to be included in the sample.
\label{fig:teinest_gamma}
}
\end{figure}

The root of this residual discrepancy is observational errors.
Not only the selection in $\teinest$ tends to favour galaxies with intrinsically larger velocity dispersion, but also those for which $\sigmaap$ has been systematically overestimated.
The probability of a detected lens being included in the SLACS sample increases with $\teinest$. This means that, at fixed lens and source redshift and velocity dispersion, lenses with a larger $\sigmaapobs$ are more likely to be selected. Due to the correlation between $\sigmaap$ and $\gamma$, such lenses have also a larger observed value of the density slope.
\Fref{fig:teinest_gamma} illustrates this effect. The top panel shows the values of $\teinest$ and $\gamma$ of the SLACS lenses that are allowed by the velocity dispersion measurements. The bottom panel shows the inferred $\pfind$ distribution, which describes the probability of a detected lens to be included in the SLACS sample, and which depends on $\teinest$. 
The uncertainties on the velocity dispersion translate into relatively large errors on $\teinest$, compared to the scale where $\pfind$ varies. Therefore, observational errors can play an important role in shaping the lens sample, particularly at low values of $\teinest$.

\begin{figure}
\includegraphics[width=\columnwidth]{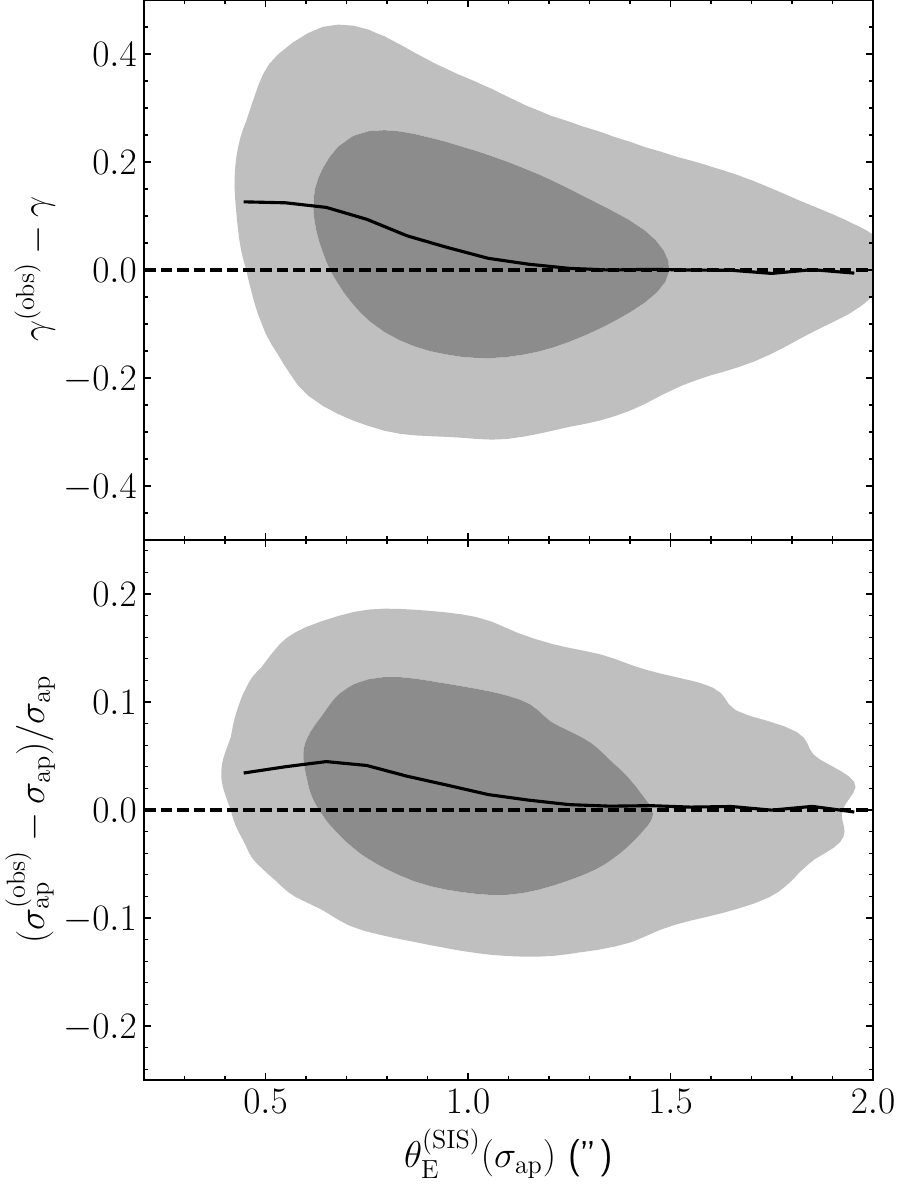}
\caption{
Bias on velocity dispersion and density slope of SLACS-like lenses.
Top: error on $\gamma$ as a function of $\teinsis$, defined in \Eref{eq:teinsis}, of posterior predicted samples of lenses, marginalised over the values of the model parameters $\hyperpars$.
Bottom: relative error on $\sigmaap$.
The solid line indicates the average in bins of $\teinsis$.
\label{fig:teinsis_bias}
}
\end{figure}

To verify this conjecture, we carried out the following test.
Given a posterior predicted sample of SLACS-like lenses, we considered the estimated Einstein radius under the assumption of a singular isothermal profile and perfect knowledge of the stellar velocity dispersion:
\begin{equation}\label{eq:teinsis}
\teinsis(\sigmaap) = 4\pi\left(\frac{\sigmaap}{c}\right)^2\frac{D_{\mathrm{ds}}}{D_{\mathrm{s}}}
\end{equation}
(the difference with \Eref{eq:one} is in the value of the velocity dispersion that is used).
Then, we looked at the relative error on $\sigmaap$ and the error on $\gamma$ as a function of $\teinsis$.
These are shown in \Fref{fig:teinsis_bias}, marginalised over the values of the parameters $\hyperpars$.
As expected, both quantities are positively biased at small values of $\teinsis$.
The average posterior predicted observed $\gamma$ is, on average (i.e. when marginalising over the model parameters), $2.10$, which is larger than the true value by $0.04$. This bias disappears for the population of detectable lenses.
In conclusion, SLACS lenses have on average a steeper density slope than their parent sample, and the value of $\gamma$ of some of their lenses has likely been overestimated.

Observational errors aside, SLACS lenses have on average a larger velocity dispersion than parent sample galaxies of the same stellar mass, as can be seen in the bottom panel of \Fref{fig:trends}. The difference is most obvious at stellar masses $m_* < 11.5$.
This is again due to the $\teinest$-based selection, because the posterior predicted distribution in $\sigmaap$ of all detectable lenses (black lines in \Fref{fig:trends}) is instead very close to that of the parent population.
Since the velocity dispersion is related to both stellar mass and size through the fundamental plane relation, it is interesting to determine whether the SLACS lenses lie on the fundamental plane or not.
If they do, then their larger $\sigmaap$ must be due to them being more compact than the average at fixed stellar mass. If not, then their values of $\sigmaap$ are genuinely larger for their stellar mass and size.
We found the latter to be the main reason.

\begin{figure}
\begin{tabular}{c}
\includegraphics[width=\columnwidth]{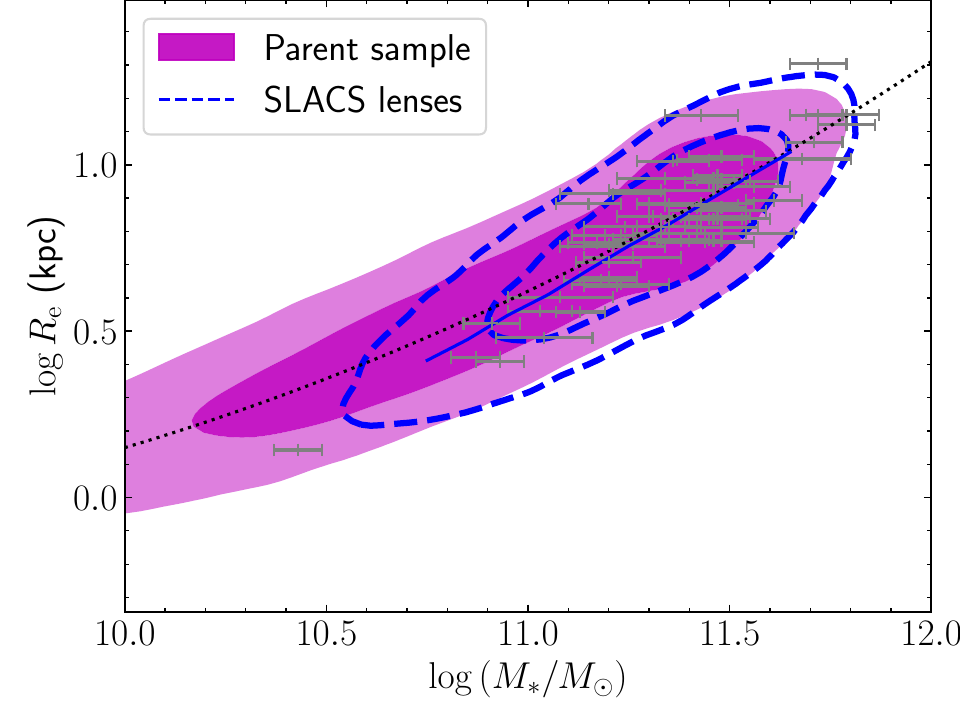} \\
\includegraphics[width=\columnwidth]{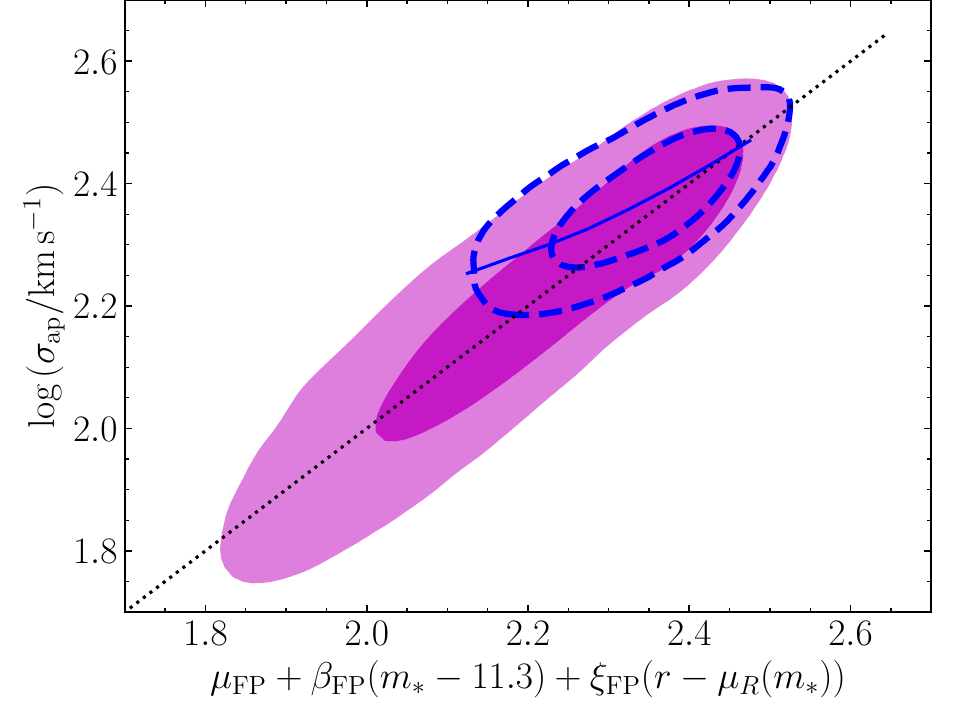} 
\end{tabular}
\caption{
Posterior predicted mass-size and fundamental plane relation.
Top: distribution in stellar mass and half-light radius of the parent population (magenta regions) and the SLACS lenses (dashed contours), marginalised over the model parameters $\hyperpars$.
The dotted black line is the mass-size relation from \citet{H+B09}.
The solid blue line indicates the average size of SLACS lenses as a function of stellar mass.
Bottom: distribution in velocity dispersion, as a function of the predicted average given the stellar mass and size (the mean of the distribution in \Eref{eq:fpmodel}). 
The solid blue line indicates the average velocity dispersion of the SLACS lenses.
\label{fig:fprel}
}
\end{figure}

To illustrate this point, we show in \Fref{fig:fprel} the posterior predicted distribution of the stellar mass-size relation (top), and the fundamental plane relation (bottom).
SLACS lenses have indeed smaller sizes than the parent population, on average, but only at the low-mass end, and only by $0.04$~dex at $m_*=11$. 
When fitting the fundamental plane relation of \Eref{eq:fpmodel} to the model for the parent population, we inferred an average value of the coefficient describing the scaling of $\sigmaap$ with excess size of $\xi_{\mathrm{FP}} \approx -0.20$. This means that a $0.04$~dex offset in size translates into an average difference in velocity dispersion of $0.008$~dex ($2\%$), corresponding to $3.5\,{\rm km}\,{\rm s}^{-1}$ at $\sigmaap=180\,{\rm km}\,{\rm s}^{-1}$. This is much smaller than the difference in $\sigmaap$ between SLACS lenses and the parent population seen in the bottom panel of \Fref{fig:trends}.
At the bottom of \Fref{fig:fprel} we show the posterior predicted distribution of the stellar velocity dispersion, as a function of the position on the fundamental plane (the mean of the distribution of \Eref{eq:fpmodel}), marginalised over the model parameters $\hyperpars$. 
SLACS lenses lie on average above the fundamental plane: their velocity dispersion is $0.035$~dex ($8\%$) higher at values of stellar mass and size corresponding to $\sigmaap=200$~km~s$^{-1}$, and the bias increases towards lower velocity dispersions.
Thus, this is the main reason for SLACS lenses having higher values of $\sigmaap$ at fixed stellar mass.
This conclusion depends to some extent on our choice of adopting the mass-size relation of \citet{H+B09}, which lies very close to that of the SLACS lenses \citep[see also][]{Aug++10}.
For a more robust assessment of the relative position of SLACS lenses and their parent population in $(M_*,\reff,\sigmaap)$ space we need consistent measurements of the three quantities, which we do not have.
Nevertheless, the main point of this investigation is that, although the fundamental plane is a relatively tight relation, it is sufficiently thick to allow for a significant bias in velocity dispersion between the SLACS lenses and the parent population.

\begin{figure}
\includegraphics[width=\columnwidth]{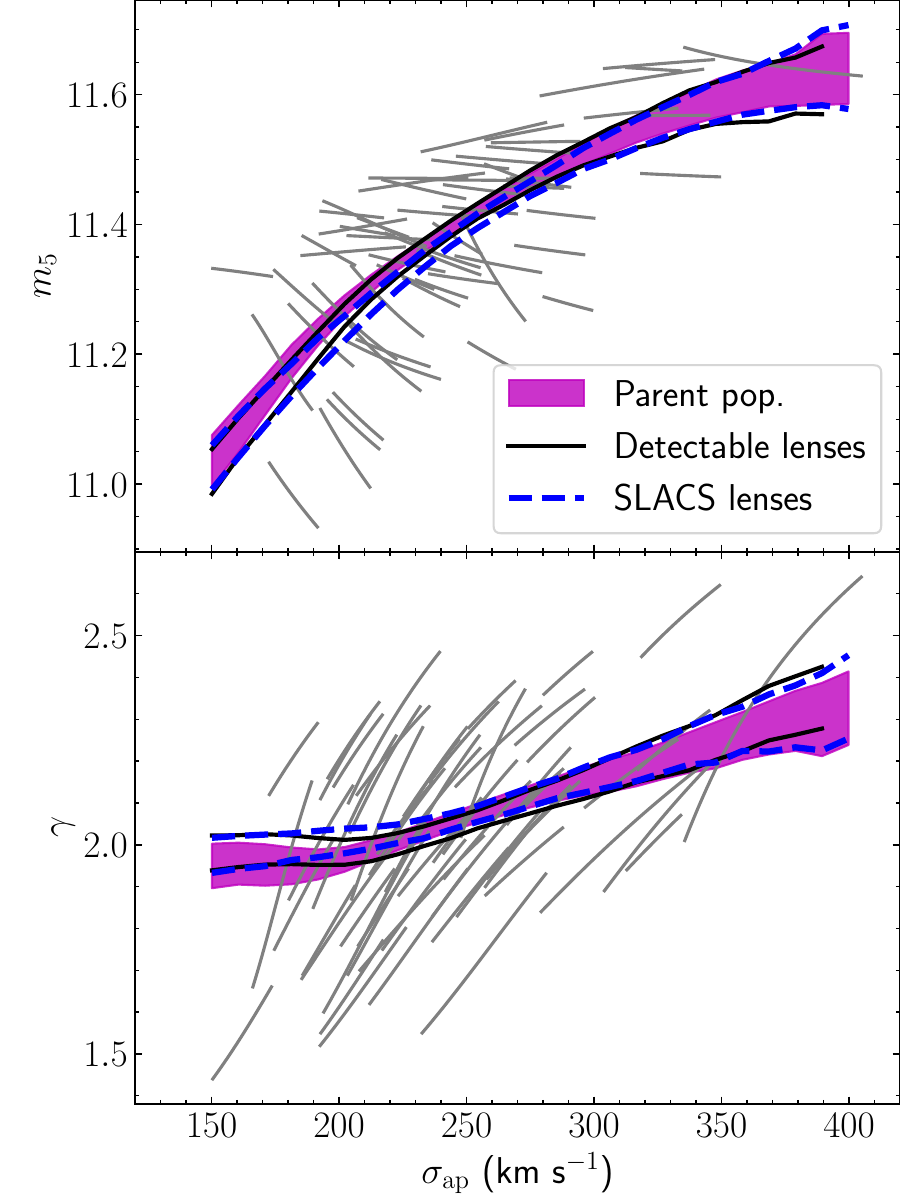}
\caption{
Posterior predicted trends with stellar velocity dispersion.
Top: average projected dark matter mass within $5$~kpc. Bottom: average density slope. 
The magenta band marks the 68\% credible region for the parent population of galaxies. The dashed blue lines enclose the 68\% credible region for the SLACS lenses. Grey lines in the top two panels span the values of the SLACS lenses that are allowed with 68\% probability.
\label{fig:sigmatrends}
}
\end{figure}

Another interesting question, related to the point above, is to what extent SLACS lenses are similar to galaxies from their parent population, once controlling for velocity dispersion.
In \Fref{fig:sigmatrends} we show the posterior predicted average true values of $m_5$ and $\gamma$ of parent sample galaxies, SLACS lenses and detectable lenses, measured in bins of velocity dispersion.
Strong lenses are remarkably similar to the parent population from this point of view: for instance, the average $m_5$ of SLACS lenses at $\sigmaap=250\,{\rm km}\,{\rm s}^{-1}$ is smaller than that of the parent sample by only $0.02$. Differences in $\gamma$ are for the most part smaller than $0.01$.
These results corroborate previous claims that SLACS lenses are representative of galaxies of the same stellar velocity dispersion \citet{Tre++06, Tre++09}, despite occupying a different region in $(M_*,\reff,\sigmaap)$ space.
However, as shown previously in \Fref{fig:teinsis_bias}, SLACS lenses have slightly biased velocity dispersion measurements. This bias needs to be taken into account when comparing them to other samples of galaxies.

\begin{figure}
\includegraphics[width=\columnwidth]{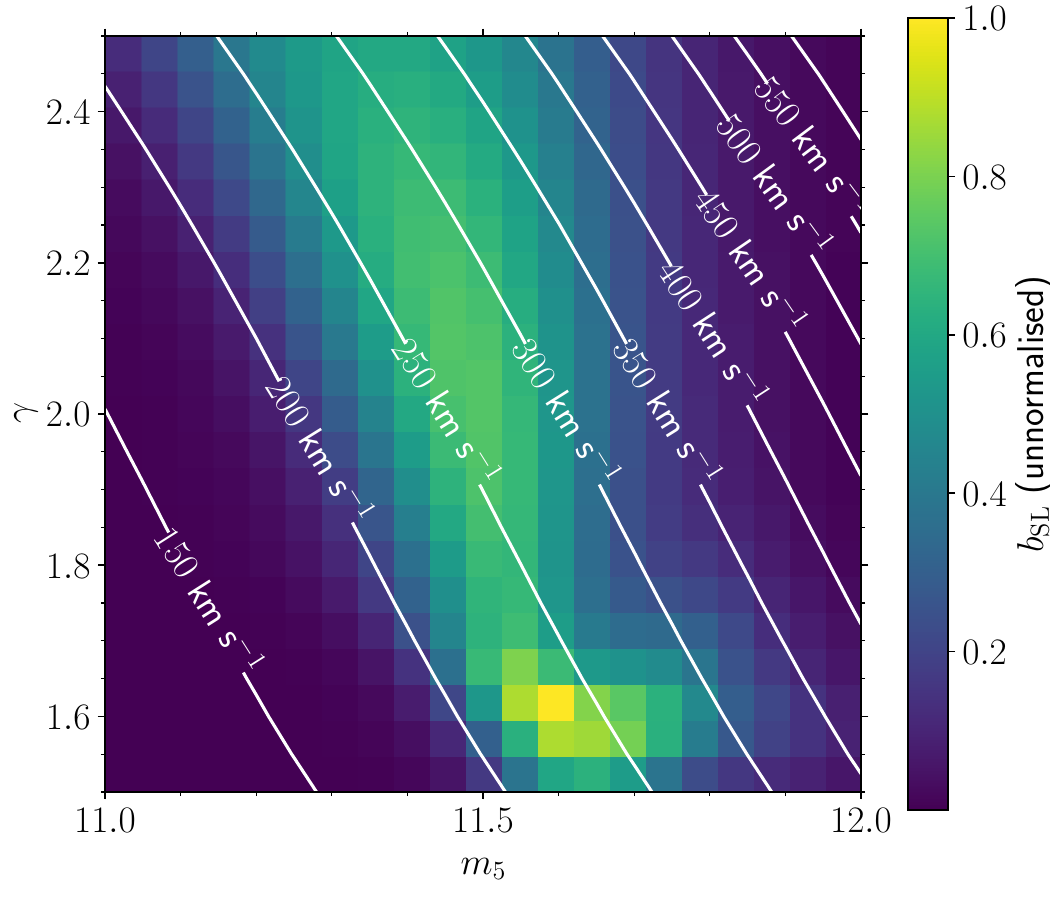}
\caption{
SLACS Strong lensing bias $\slbias$ for a galaxy at $z=0.2$ with $\reff=7$~kpc, as a function of $m_5$ and $\gamma$.
White lines connect points of constant velocity dispersion, with $\sigmaap$ equal to the value indicated.
We computed $\slbias$ from \Eref{eq:simpleslbias}, while marginalising over the model parameters $\hyperpars$.
\label{fig:slbias}
}
\end{figure}

As a last thing, we computed the strong lensing bias $\slbias$ of the SLACS survey. We introduced $\slbias$ in section \ref{ssec:problem}: it is the function by which the distribution of parent sample galaxies is multiplied to produce the sample of strong lens galaxies observed in a survey. 
We computed $\slbias$ from \Eref{eq:simpleslbias} for an example galaxy at redshift $z=0.2$ and with half-light radius $\reff=7$~kpc, as a function of $m_5$ and $\gamma$.
\Fref{fig:slbias} shows the result, marginalised over the model parameters $\hyperpars$.
The strong lensing bias peaks on a relatively narrow band in $m_5-\gamma$ space, which can be seen as a window function of the SLACS survey: galaxies that lie on that band are more likely to be selected as SLACS lenses, while those outside of it are missed. 
In \Fref{fig:slbias} we also overplotted lines of constant velocity dispersion, which are nearly parallel to the lines of constant $\slbias$. 
This matches our view that the selection function of SLACS is mostly one in velocity dispersion: at fixed $\sigmaap$, the strong lensing bias does not vary much with either $m_5$ or $\gamma$, hence SLACS lenses have a similar density profile to regular early-type galaxies with the same velocity dispersion.

The expression for the strong lensing bias is independent of the distribution of the parent galaxy population $\prlens$.
One could then consider a different parent sample from the one used by SLACS, and use $\slbias$ to predict what kind of lenses would be discovered in a SLACS-like campaign with this different sample. In general this could be a valid use of $\slbias$, but not in our case. This is because our model for the lens finding probability, which is meant to describe the prioritisation of targets for follow-up observations, depends explicitly on the observed velocity dispersion. But in a hypothetical sample with a different distribution in $\sigmaap$, a SLACS-like prioritisation that selects a fixed number of targets at the top of the $\teinest$ distribution would correspond to a different selection in $\sigmaap$, and consequently different $\pfind$. The question of what a SLACS-like survey would look like with a different parent population is then an ill-posed one: more details on the follow-up strategy need to be provided.

\section{Discussion}\label{sect:discuss}

Joint lensing and dynamics studies of the SLACS lenses carried out so far have found that the average density slope of the 59 lenses in the sample is $\left<\gamma\right>\approx2.1$.
According to our analysis, this value of the average $\gamma$ is larger than that of regular early-type galaxies of the same stellar mass, by about $0.1$.
This result can have important implications for studies of the evolution in the density profile of massive early-type galaxies.

\citet{Ruf++11}, \citet{Bol++12} and \citet{Son++13b} showed how, when combining SLACS with other samples of lenses, the average $\gamma$ tends to be larger at lower redshift, pointing towards a steepening of the total density profile with time.
The common feature among these studies is that they all rely on the SLACS sample as a low-redshift pivot. Thus, if $\left<\gamma\right>$ of SLACS were to be lowered, the conclusions would change dramatically. 
For instance, \citet{Son++13b} measured $d\left<\gamma\right>/dz = -0.31\pm0.10$ when combining SLACS with SL2S lenses. The average redshift of the SL2S sample is $\left<z\right>\approx0.5$, larger than that of SLACS by $\approx0.3$. This means that a systematic offset in $\left<\gamma\right>$ of the SLACS lenses by $0.1$ could account for the detected trend entirely.
A similar argument can be made for the study of \citet{Bol++12}, which is based on the BELLS sample. 
BELLS lenses were discovered spectroscopically with a very similar method employed for SLACS, including a criterion for the prioritisation of photometric follow-up observations based on the estimated Einstein radius $\teinest$.
Crucially, however, BELLS did not rely on the observed velocity dispersion in the estimate of $\teinest$, due to the low signal-to-noise ratio of the available spectroscopic measurements: instead, they assigned a fixed value of $\sigmaap=250\,{\rm km}\,{\rm s}^{-1}$ to all lens candidates \citep{Bro++12}.
As a result, the BELLS sample is immune from the selection in velocity dispersion that causes the bias in $\gamma$ of the SLACS lenses.

In light of these findings, the evidence for a steepening with time of the total density profile of massive early-type galaxies needs to be revisited.
SLACS lenses cannot be simply compared to lenses at higher redshift from different samples, but their velocity dispersion selection needs to be taken into account. It is possible that the value of $d\left<\gamma\right>/dz$ will be revised upwards after correcting for selection effects. That would help reconcile lensing and dynamics measurements with some lensing-only studies \citep{Tan++24, Sah++24} and with theoretical predictions \citep{Rem++13, SNT14, Xu++17, Sha++18}.
Nevertheless, in order to know with certainty whether the steepening in $\gamma$ is real or not, it is necessary to correct for selection effects also the other samples of lenses mentioned above. That, however, is beyond the scope of this work.

Our findings have implications for time-delay cosmography studies as well.
\citet{Bir++20} used the SLACS sample to put a prior on the lens mass model of time-delay lenses from the TDCOSMO sample \citep{Mil++20}, under the assumption that any differences between the SLACS and the TDCOSMO sample can be accounted for by scalings with the ratio between Einstein radius and half-light radius.
Given the peculiar selection function of SLACS lenses, which tends to pick objects that are offset from the fundamental plane of early-type galaxies, it seems unlikely that the two samples can be combined in such a way without introducing biases.
However, we are unable to quantify such a bias, which might well be within the statistical uncertainties, because the mass model used by \citet{Bir++20} is more complex than the simple power-law on which our study is based.

All of our analysis was carried out under the assumption of a spherical mass distribution and isotropic orbits for the stars. 
This assumption allowed us to convert directly the Einstein radius and stellar velocity dispersion of each lens into an inference on the two free parameters of the power-law mass profile, $(m_5,\gamma)$.
Although the amount of flattening and anisotropy of the SLACS lenses has been found to be modest \citep{Bar++11}, high precision measurements would require us to allow for a departure from spherical symmetry and from the isotropy assumption.
At fixed mass and tracer density profile, radial orbits tend to increase the stellar velocity dispersion.
Fitting an isotropic power-law model to a radially anisotropic lens galaxy would then lead to an overestimate of the density slope $\gamma$ \citep[see][ for a quantification of this effect]{Koo++06}.
We expect the orbital anisotropy to vary from galaxy to galaxy: this is a source of scatter in the velocity dispersion at fixed mass profile that we are absorbing into the inferred distribution in $\gamma$.
As a result, we likely overestimated the strength of the correlation between $\gamma$ and the velocity dispersion.
Because, as we established, the nature of the bias in $\gamma$ between the SLACS lenses and the parent population is entirely due to the selection in velocity dispersion, an intrinsic scatter in orbital anisotropy would mean that what we interpreted exclusively as a bias in $\gamma$ is partly a bias in anisotropy: 
SLACS lenses are not as biased in $\gamma$ with respect to the parent population, but have on average more radially anisotropic orbits.
A similar argument can be made regarding the three-dimensional structure.
This does not change the main point of our analysis, which is that the way that SLACS galaxies were selected makes them different from galaxies with similar stellar mass and size, in terms of their internal structure.

One of the limiting factors in our analysis is the lack of consistent measurements of the stellar mass profile of galaxies of the SLACS lenses and the parent population galaxies. 
This prevented us from taking full advantage of the information available on the parent sample, and forced us to adopt the more relaxed prior on the fundamental plane described in section~\ref{ssec:fp}.
For future studies, it would be beneficial to have access to the same photometric data for both the lenses and the non-lenses.
We expect that strong lens studies based on data from Euclid \citep{Mel++24} will satisfy this requirement.


\section{Conclusions}\label{sect:concl}

We carried out a joint lensing and dynamical analysis of the SLACS sample of lenses.
Our study differed from previous analyses of the same sample \citep{Aug++10,Son++13b,Pos++15,Sha++21} under one crucial aspect: we corrected the inference for selection effects.
Our model for the selection function describes both the probability of detecting a lensed source in the spectroscopic and photometric data on which the survey was based, and the probability of a lens candidate being included in the final SLACS sample.
This constitutes a step forward in complexity and fidelity with respect to earlier attempts \citep{Son++15,Son++18}.

We constrained the distribution in the mass structure of massive early-type galaxies, under the assumption of a spherically symmetric power-law mass density profile and isotropic orbits.
We found that galaxies with a stellar mass of $\log{(M_*/M_\odot)}=11.3$ and average size for their mass have on average a total projected mass enclosed within $5$~kpc of $\log{M_5} = 11.332\pm0.013$, and an average density slope of $\gamma=1.99\pm0.03$.
By contrast, a fit of the same mass model to the SLACS lenses gives a larger average density slope by approximately $0.1$.
We identified the source of this bias to lie entirely in the way that the SLACS lens candidates were prioritised for photometric follow-up observations, which relied on the observed velocity dispersion. As a result, SLACS lenses have larger velocity dispersion than regular early-type galaxies at fixed stellar mass and size, and their velocity dispersion measurements are likely biased towards larger values.
Nevertheless, they are fairly representative of galaxies of similar velocity dispersion (though they have slightly lower masses).
We also corroborated evidence towards early-type galaxies not being homologous systems: their total mass grows more strongly with stellar mass and decreases less strongly with half-light radius than a purely homologous structure would suggest.

Because the source of the bias of the SLACS lenses is in the target prioritisation, which is a peculiarity of SLACS, we do not expect other lens samples to be similarly biased. This could have implications for previous claims of an evolution in the total density profile of massive early-type galaxies that relied on the combination of SLACS data with other lens samples.

Our study provides a reference on which to base future investigations that require knowledge of the selection function of a strong lens sample.
One immediate use case is testing the predictions of theoretical models on the mass structure of early-type galaxies: this can be done directly by taking the model distribution in $(m_5,\gamma)$ that we inferred from the data and comparing it with theory.
If the goal is building a large sample of galaxies with similar internal structure as the SLACS lenses, for instance to carry out weak lensing measurements \citep[such as in the study of][]{Son++18}, this can be done by matching the two samples in velocity dispersion, provided that differences in mass of $0.02$~dex are deemed not important.
If one wishes to use information from the SLACS sample as a prior for different samples of lenses, as done in the time-delay lensing analysis of \citet{Bir++20}, additional work is needed: the strong lensing bias of each strong lens sample needs to be determined separately. 
Finally, by applying the method developed in this paper to more complex models and different samples of lenses, we can remove the strong lensing bias on other properties of the mass structure of galaxies, such as the stellar mass-to-light ratio, the dark matter distribution, and their time evolution.



\begin{acknowledgements}

This work was initiated at the 2023 KICP workshop “Lensing at different scales: strong, weak, and synergies between the two”.
I thank Tommaso Treu, Tom Collett, Tian Li, Anowar Shajib, Natalie Hogg and Birendra Dhanasingham for their useful comments and suggestions.

\end{acknowledgements}

\bibliographystyle{aa}
\bibliography{references}

\end{document}